\documentclass[printer]{aa}
\usepackage{natbib}
\bibpunct{(}{)}{;}{a}{}{,} 
\usepackage{graphicx}
\usepackage{txfonts}
\usepackage{color}

\begin{document}
\title{Hot-spot model for accretion disc variability as random process -- II.\\ Mathematics of the power-spectrum break frequency}
\author{Tom\'a\v{s}~Pech\'a\v{c}ek,\inst{1} Ren\'e~W.~Goosmann,\inst{2} Vladim\'{\i}r~Karas,\inst{1} Bo\.{z}ena~Czerny,\inst{3} \and  Michal~Dov\v{c}iak\inst{1}}
\institute{Astronomical Institute, Academy of Sciences, Bo\v{c}n\'{\i}~II 1401, CZ-14131~Prague, Czech Republic 
\and Observatoire Astronomique de Strasbourg, F-67000 Strasbourg, France
\and Copernicus Astronomical Center, Bartycka 18, P-00716~Warsaw, Poland}

\authorrunning{T.~Pech\'a\v{c}ek, R. W. Goosmann, V. Karas, B. Czerny, \& M. Dov\v{c}iak}
\titlerunning{Accretion disc variability as random process}
\date{Received 6 September 2012; accepted 10 June 2013}
\abstract{We study some general properties of accretion disc variability in the context of stationary random processes. 
In particular, we are interested in mathematical constraints that can be imposed on the functional form of the Fourier 
power-spectrum density (PSD) that exhibits a multiply broken shape and several local maxima.}{We develop a methodology 
for determining the regions of the model parameter space that can in principle reproduce a PSD shape with a given number 
and position of local peaks and breaks of the PSD slope. Given the vast space of possible parameters, it is an important 
requirement that the method is fast in estimating the PSD shape for a given parameter set of the model.}
{We generated and discuss the theoretical PSD profiles of a shot-noise-type random process with exponentially decaying flares. 
Then we determined conditions under which one, two, or more breaks or local maxima occur in the PSD. We calculated positions 
of these features and determined the changing slope of the model PSD. Furthermore, we considered the influence of the modulation 
by the orbital motion for a variability pattern assumed to result from an orbiting-spot model.}
{We suggest that our general methodology can be useful in for describing non-monotonic PSD profiles 
(such as the trend seen, on different scales, in exemplary cases of the high-mass X-ray binary Cygnus X-1 and the 
narrow-line Seyfert galaxy Ark 564). We adopt a model where these power  spectra are reproduced as a superposition of
 several Lorentzians with varying amplitudes in the X-ray-band light curve. Our general approach can help in constraining 
 the model parameters and in determining which parts of the parameter space are accessible under various circumstances.}
{} 
\maketitle

\section{Introduction}

Accretion discs are believed to drive the variability of supermassive black holes in 
active galactic nuclei (AGN) and, on much smaller time-scales, also the fluctuating signal from Galactic black holes
in accreting compact binaries \citep{1992ApJ...391..518V,1999ApJ...510..874N,2006Natur.444..730M}. 
Observed light curves, $f\,\equiv\,f(t)$, exhibit irregular, featureless variations at all frequencies that 
can be studied \citep{1987Natur.325..694L,1987Natur.325..696M,2006ASPC..360.....G}.  Nonetheless,
accretion is the common agent \citep{2002apa..book.....F}. 

Simplified models of accretion disc processes do not explain {\em{}every} aspect of variability
of different sources. Phenomenological studies also show a considerable complexity of the
structure of accretion flows. Accretion discs co-exist with their coronae and possibly winds 
and jets, which produce a highly unsteady contribution, and so the observed signal of the source is determined 
by the mutual interaction of different components \citep{2007A&ARv..15....1D}. 
In this paper we study X-ray variability features that can be attributed to stochastic
flare events that are connected to the accretion disc. Our aim is to constrain the resulting variability using a 
very general scheme. 

Random processes provide a suitable framework for a mathematical description of the measurements of 
a physical quantity that is subject to non-deterministic evolution, for example due to noise 
disturbances or because of the intrinsic nature of the underlying mechanism. In our previous paper 
\citep[][Paper~I]{2008A&A...487..815P} we studied the theory of random processes as a tool for 
describing the fluctuating signal from accreting sources, such as AGN 
and Galactic black holes observed in X-rays. These objects exhibit a featureless variability on 
different time-scales, probably originating from an accretion disc surface and its corona. 
In particular, we explored a scenario where the expected signal is generated by an ensemble 
of spots randomly created on the accretion disc surface, orbiting with Keplerian velocity 
at each corresponding radius \citep[][and references therein]{1979ApJ...229..318G,2004A&A...420....1C}.

Various characteristics can be constructed from the light curve of an astronomical object
to study its variability properties \citep{1992scma.book.....F}. The
standard approach is to analyse the light curve in terms of its power spectral density (PSD), 
which provides the variability power as a function of frequency \citep{2008arXiv0802.0391V}.
The histogram of the flux distribution measures the time-scale that the source spends at a given flux level. 
Furthermore, the root-mean-square (rms) scatter can be computed for different sections
of the light curve that correspond to different flux levels. This establishes the rms-flux relation
\citep[]{2001MNRAS.323L..26U}.

In this context an important reference to our work is given by \citet{Uttley2005}, where the authors presented a model
to explain the observed linear rms-flux relations that are derived from X-ray observations of accreting black hole
systems. The model generates light curves from an exponential applied to a Gaussian noise, which by construction produces 
a log-normal flux distribution consistent with the observed X-ray flux histograms of the Galactic black hole
binary Cyg-X1, as well as of other sources. \citet{Uttley2005} concluded that the observation of a log-normal flux distribution
and the linear rms-flux relationship imply that the underlying variability process cannot be additive. 
Indeed, this statement holds for a large family of shot-noise variability models and it was extended to other
sources in different spectral states \citep{2006MNRAS.367..801A}, also in the optical band 
\citep{2009ApJ...697L.167G}. 

As mentioned above, in this paper we concentrate on characteristics of the PSD profile, in particular 
on the constraints that can be derived from the PSD slope and the occurrence of break frequencies. 
Because power spectra do not provide complete information, even a perfect agreement between 
the predicted PSD and the data cannot be considered as a definitive confirmation of the 
scenario. We do not study additional constraints, such as the rms-flux relation or time lags 
between different bands, which certainly have a great potential of distinguishing 
between different schemes. Indeed, these characteristics provide complementary
pieces of information that may be found incompatible with the spot model. In other words, the model
presented here is testable, at least in principle, and it will be possible to confirm the model 
or reject it by observations.

Although well-known ambiguities are inherent to Fourier power spectra, an ongoing 
debate is held about the shape of the PSD and the dominant features that 
could reveal important information about the origin of the variability. The broad-band 
PSD has been widely employed to examine the fluctuating X-ray light curves, often 
showing a tendency towards flattening at low frequencies \citep{1993ApJ...414L..85L,
1993ARA&A..31..717M,2002MNRAS.332..231U}. PSDs of interest for us  
are characterized by a power-law form of the Fourier power spectrum $\omega S(\omega)$, which 
decays at the high-frequency part proportionally to the first power of $\omega$. A break occurs at
lower frequencies, where the slope of the power-law changes. The profile 
of the power spectrum can be even more complex, showing multiple breaks or even local peaks 
separated by minima of the PSD curve.

The occurrence of a break frequency has been discussed by various authors 
\citep[e.g.][]{1999ApJ...510..874N,2003ApJ...593...96M} for its obvious relevance for understanding
observed PSDs and interpreting the underlying mechanism that causes the slope change. 
In this context, \citet{2005MNRAS.359.1469M} have examined the case of the Seyfert 1 galaxy 
MCG--6-30-15 with RXTE and XMM-Newton data. Very accurate fits suggest a multi-Lorentzian 
structure instead of a simple power-law. Furthermore, \citet{2007MNRAS.382..985M}, combining the 
XMM-Newton, RXTE and ASCA observations, discovered multiple Lorentzian components in the X-ray 
timing properties extending over almost seven decades of frequency ($10^{-8}\lesssim f\lesssim10^{-2}$~Hz) 
of the narrow-line Seyfert~1 galaxy Ark 564. These authors concluded that the evidence points to two 
discrete, localized regions as the origin of most of the variability. \citet{2011ApJ...743L..12M} 
discussed the AGN optical light curves monitoring by Kepler. They found, rather surprisingly, that the
PSD exhibits power-law slopes of $-2.6$ to $-3.3$, i.e., significantly steeper than typically seen in the X-rays. 
Furthermore, within the category of stellar-mass black holes, \citet{2003A&A...407.1039P} analysed 
Cygnus X-1 PSD from about three years of RXTE monitoring. These authors concluded that the changing 
form of the power spectrum can be interpreted as a combination of Lorentzians with gradually 
evolving amplitudes.

In this paper, motivated by this and similar observational evidence, 
we embark on a mathematical study of PSD that can be fitted either with a broken 
power-law or a double-bending 
power law, with the variability power $\omega\,S(\omega)$ dropping at low and high frequencies, 
respectively. We consider a simplified (but still non-trivial) description of the intrinsic 
variability (random flares with idealized light curve profiles of the individual events), 
which allows us to develop an analytical approach to the resulting observable PSD profile. 
In particular, we determine conditions under which one, two, or more breaks or 
local maxima occur in the PSD. We calculate positions of these features and determine the 
changing slope of the model PSD.

The paper is organised as follows. In section \ref{sec1} we discuss the PSDs generated by a shot-noise-like 
random process with exponentially decaying flares. We developed a general approach to describe the morphology 
of the power spectra and demonstrate the results on a simple, analytically solvable example of a bi-Lorentzian PSD. 
In section \ref{secDoppler} we consider the influence of the modulation by the orbital motion on the PSDs. 
We summarise our conclusions and give perspectives for the application of this fast, analytical modelling scheme 
in section~\ref{conclusions}.

\section{Exponential flares}
\label{sec1}
It is well known that the power spectra of stationary random processes
must be flat at zero ($S(\omega)\propto\omega^0$) and integrable over
all frequencies  (i.e. decaying faster than $\omega^{-1}$ for large
$\omega$). Moreover,  it was demonstrated for the multi-flare model 
\citep[]{2008A&A...487..815P} that the high-frequency limit of the PSD is a power-law,
$S(\omega)\propto\omega^\gamma$,  with the slope $\gamma<-1$
determined by the intrinsic shape of the flare-emission profile. The
low frequency limit depends mainly on the statistics of  the 
flare-generating process. The form of the intermediate part of the power
spectrum   is influenced by the interplay of all of these effects. The
simplest possible power  spectra generated by the flare model are
approximately of the broken power-law form.

It is natural to assume that the power spectrum break-frequency is
associated with  some intrinsic time-scale of the process. This
heuristic approach is supported  by the basic scaling properties of
the Fourier transform: rescaling the process in temporal domain by a
factor $a$ shifts the break frequency by $a^{-1}$. The power-law PSDs
are invariant under this rescaling, and the break frequency is the only
feature of the power spectra whose position changes. However, we will
see  that for two reasons it is difficult to go beyond this trivial
observation. 

Rigorous definition of the break frequency can be a problem. Even for the simplest
acceptable feature-less power spectra such as a Lorentzian PSD, which are power-laws 
to high accuracy for most of frequencies, the change of slope occurs smoothly over a long interval.
It is therefore not obvious which frequency should be defined as the break frequency, and even
if we decide on some particular definition, various problems of interpretation may emerge.

In this introductory section, we constrain our discussion to mutually independent 
exponentially decaying flares with emissivity profiles $I(t,\tau)=I_0(\tau)\exp(-t/\tau)\,\theta(t)$,
where $\theta(t)$ is the Heaviside function and $\tau$ is the lifetime of individual flares.
Exponentials are well suited for our investigations because of their simple, feature-less power  spectra.
Moreover, there is a time-scale $\tau$ naturally and uniquely associated with each flare.
This allows us to relate the break frequencies to the ensemble averages over the range of 
$\tau$.

To simplify the problem even more, we do not take into account the relativistic
effects in the rest of this section. The power spectrum of a random 
shot-noise-like process consisting of mutually independent exponentially decaying flares is given by 
\begin{equation}
S(\omega)=\lambda\int \frac{\tau^2}{1+\tau^2\omega^2}\, I^2_0(\tau)\, p(\tau)\,{\rm d}\tau,
\label{GenSpec}
\end{equation}
where $\lambda$ is the mean rate of flares and $p(\tau)$ is the probability distribution 
of $\tau$. In the case of identical emissivity profiles, i.e. with 
$p_{\rm L}(\tau)=\delta(\tau-\tau_0)$, the power spectrum is a pure Lorentzian,
\begin{equation}
S_{\rm L}(\omega)=\frac{\lambda\, I^2_0\,(\tau_0)\,\tau_0^2}{1+\tau_0^2\omega^2}.
\label{Lorentzian}
\end{equation} 
Even this simple model can reproduce a wide variety of power spectra. In the context of the variability
of the X-ray sources the model was introduced by \citet[]{1989ESASP.296..499L}.

It is traditional to plot the PSDs in $\omega$ versus $\omega S(\omega)$ graphs.
It can be easily shown that the function $\omega S_{\rm L}(\omega)$ reaches
its maximum at $\omega_{\rm M}=\tau_0^{-1}$, as one can intuitively expect.
In the general case the position of this maximum depends on the actual shape of 
the functions $p(\tau)$ and $I_0(\tau)$. Since the power spectrum is a nonlinear
functional of the signal, the peak frequency differs from both ${\rm E}\left[\tau\right]^{-1}$ and
${\rm E}\left[\tau^{-1}\right]$, where ${\rm E}\left[.\right]$ denotes the averaging operator.

To demonstrate this nonlinear behaviour we will next study a process that
consists of exponentials with only two different time-scales, i.e. a process with 
$p(\tau)=A\delta(\tau-\tau_1)+(1-A)\delta(\tau-\tau_2)$, where $A$ is some constant
from the interval $\langle 0,1\rangle$ and $\tau_1>\tau_2$ . The resulting power spectrum is 
\begin{equation}
S_{2{\rm L}}(\omega)=\frac{\lambda \, A\, I^2_0(\tau_1)\, \tau_1^2}{1+\tau_1^2\omega^2}
+\frac{\lambda\, (1-A)\, I^2_0(\tau_2)\, \tau_2^2}{1+\tau_2^2\omega^2}.
\label{2Lorentziany}
\end{equation} 
As long as we are interested only in the process times-cales, the PSD normalization is irrelevant.
Without loss of generality we can rescale the power spectrum both in normalization and in frequency domain as 
\begin{equation}
S_{\rm r}(\omega)=[S_{2{\rm L}}(0)]^{-1}S_{2{\rm L}}(\tau_1^{-1}\omega)=\frac{\alpha}{1+\omega^2}
+\frac{1-\alpha}{1+K^2\, \omega^2}.
\label{2Rorentziany}
\end{equation}
Both $K$ and $\alpha$ are from the interval $\langle 0,1\rangle$.   
The peak frequency $\omega_{\rm M}$ is given by the equation
\begin{equation}
\left.\frac{\rm d}{{\rm d}\omega}\left[\omega\, S_{\rm r}(\omega)\right]
\right|_{\omega=\omega_{\rm M}}=0,
\end{equation}
which leads to the bicubical equation 
\begin{eqnarray}
-K^2\left[(1-\alpha)+\alpha K^2\right]x^3+\left[(1-\alpha)
-K^2(2-\alpha K^2)\right]x^2\nonumber \\
+\left[(1-\alpha)(2-K^2)+\alpha (2K^2-1)\right]x+1=0,
\label{bicub}
\end{eqnarray}
where $x=\omega^2_{\rm M}$. 

\begin{figure*}[tbh]
\includegraphics[width=0.49\textwidth]{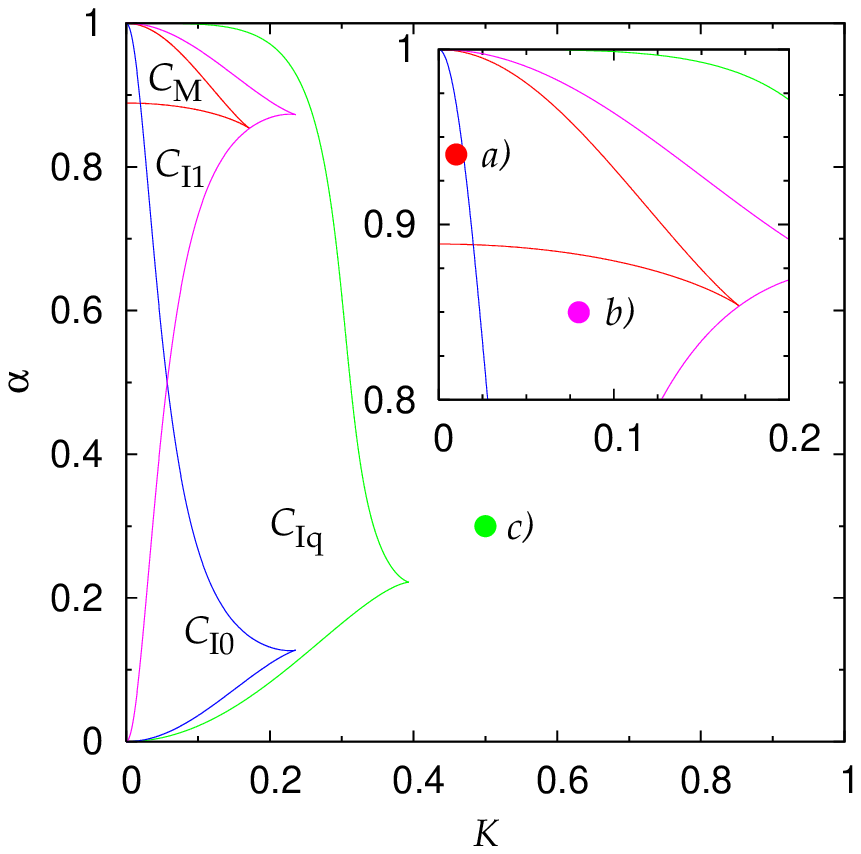}
\includegraphics[width=0.49\textwidth]{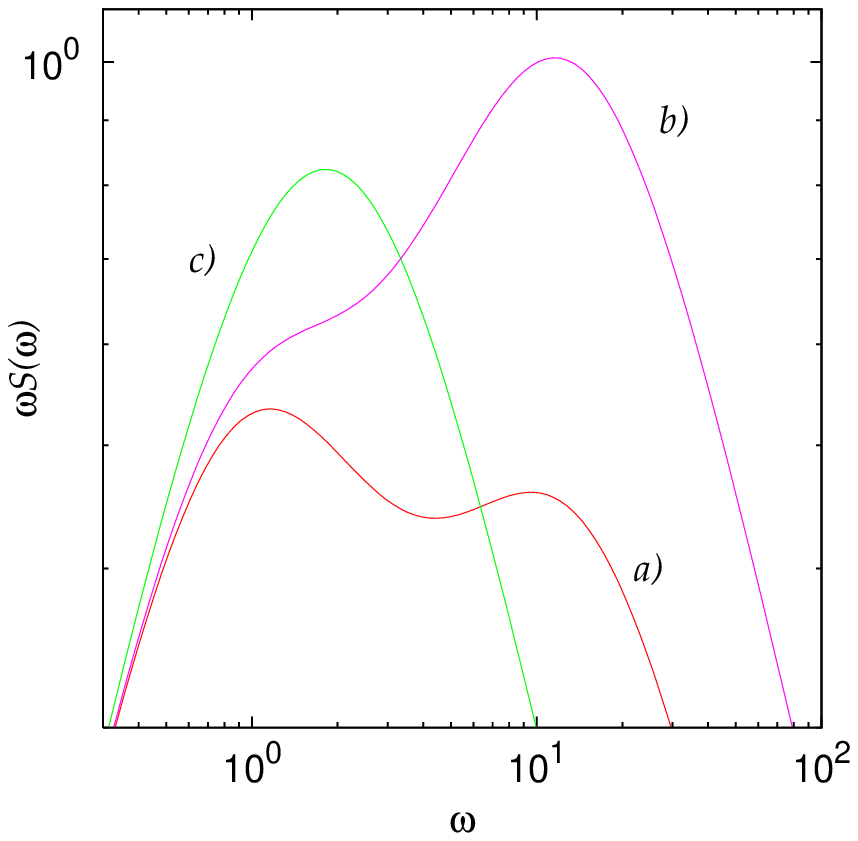}
\caption{Left panel: Diagram describing the morphology of the PSD profiles according to eq.\ (\ref{2Rorentziany}) 
for all possible values of the parameters $\alpha$ and $K$. Different curves represent the boundaries of subsets
of the parameters for which the PSD exhibits a double-feature. The red curve encloses
set $C_{\rm M}$ of the doubly peaked power spectra. The other curves correspond to the sets of power spectra
with a nontrivial structure of inflection points $C_{\rm I0}$ (blue), $C_{\rm I1}$ (magenta) and
$C_{\rm Iq}$ (green). Right panel: Three examples of the PSD profiles. A double-peaked power spectrum (red)
with parameters $\alpha=0.94$ and $K=0.01$ (i.e., the values within set $C_{\rm M}$, as denoted by 
the red point in the left panel), doubly-broken
power spectrum with a single local maximum (magenta) corresponding to $\alpha=0.85$ and $K=0.08$ 
(the magenta point within set $C_{\rm I1}$ and outside of $C_{\rm M}$) and a power spectrum with 
single maximum and break (green) with $\alpha=0.5$ and $K=0.5$ (the green point outside of $C_{\rm Iq}$).}
\label{zobak}
\end{figure*}

\begin{figure*}[tbh]
\includegraphics[width=0.32\textwidth]{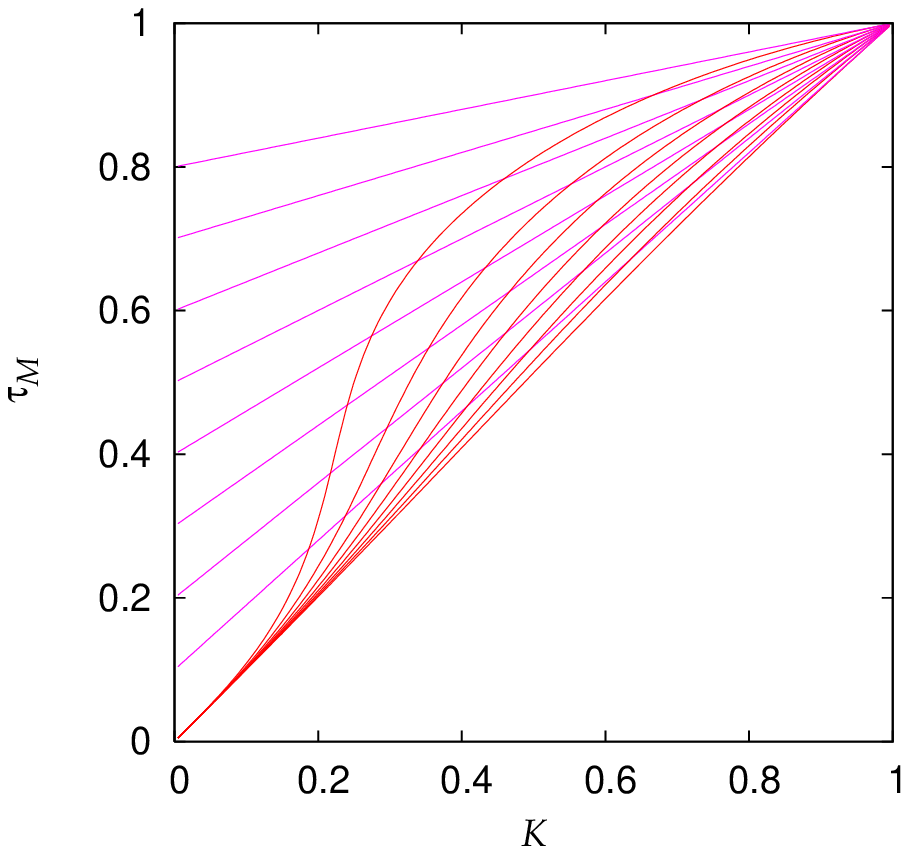}
\includegraphics[width=0.32\textwidth]{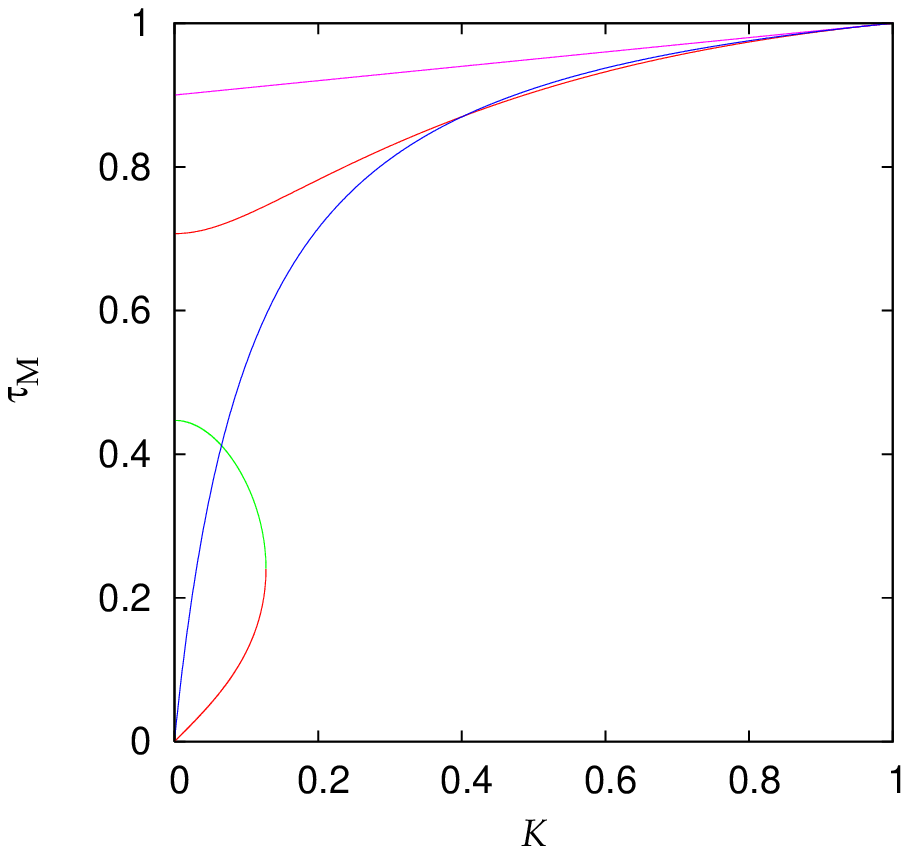}
\includegraphics[width=0.32\textwidth]{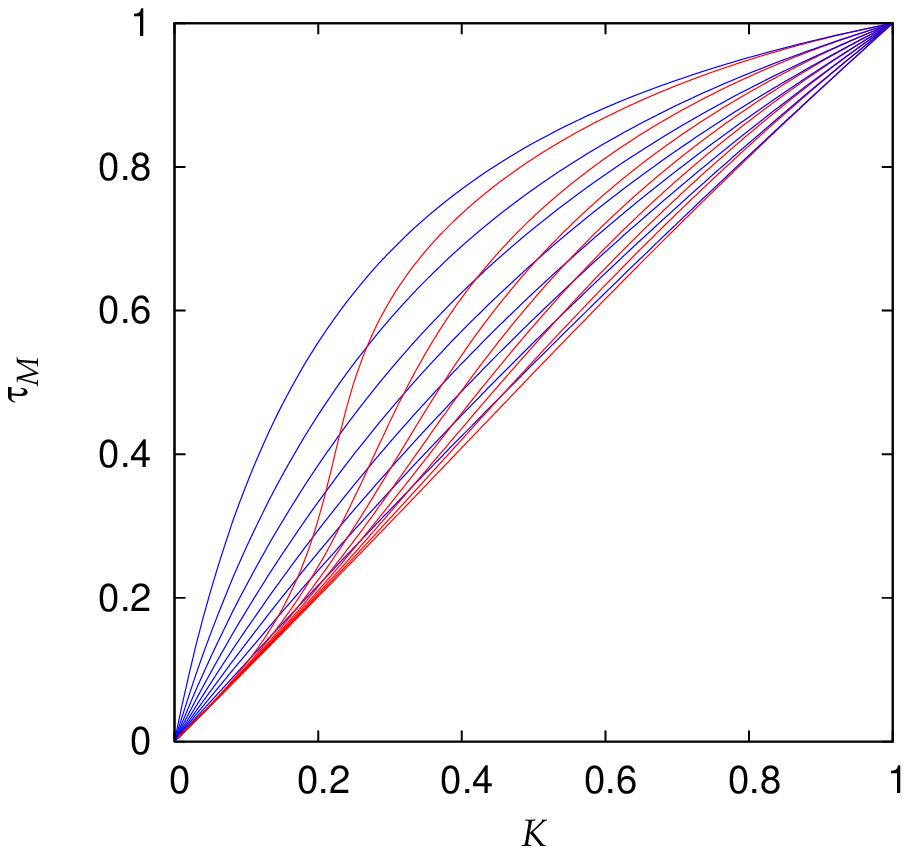}
\caption{
Left: Comparison of the time-scales $\tau_{\rm M}$ (red) and ${\rm E}\left[\tau\right]$ (magenta).
The lines are calculated for $\alpha$ from the set $\{0.1,\,0.2,\,0.3,\,0.4,\,0.5,
\,0.6,\,0.7,\,0.8\,\}$, avoiding set $C_{\rm M}$.
Middle: As in the left panel, but for $\alpha=0.9$. For low $K$ this corresponds to double-peaked
power spectra. The red curves correspond to $1$ over the peak frequency, the green is for the minimum.
The blue curve denotes the time-scale $1/{\rm E}\left[1/\tau\right]$.
Right: Comparison of the time-scales $\tau_{\rm M}$ (red) and $1/{\rm E}\left[1/\tau\right]$ (blue).}
\label{kridlo}
\end{figure*}

Equation (\ref{bicub}) can have either one or three positive real roots. For most 
of the combinations of parameters $\alpha$ and $K$ the root is unique and the corresponding power spectrum 
has a single global maximum. Three roots do occur only for parameters within a small subset of the configuration
space $C_{\rm M}$. The corresponding power spectra have two local maxima separated by a local minimum.
The boundary of set $C_{\rm M}$ is denoted by the red line in figure~\ref{zobak}.
Note that after rescaling the intrinsic time-scales of the two exponentials
are $1$ and $K$. The mean time-scale and the mean frequency of the process are then given by
\begin{eqnarray}
{\rm E}\left[\tau\right]&=&\alpha+(1-\alpha)K,\\
{\rm E}\left[\tau^{-1}\right]&=&\alpha+(1-\alpha)K^{-1}.
\end{eqnarray}
However, the break time-scale $\tau_{\rm M}=1/\omega_{\rm M}$ depends nonlinearly
on both $\alpha$ and $K$. The comparison of these three quantities is plotted in
the figure~\ref{kridlo}.  We can see that the displacement of the actual peak position
and the two linearly averaged time-scales can be very significant.

Note 
that our formulae are expressed and graphs are labelled in arbitrary (dimensionless)
units. When the underlying mechanism is interpreted in terms of flaring events in a
black hole accretion disc, the meaning of these units can be identified with 
geometrical units. Corresponding quantities in physical units can be
obtained by a simple conversion:
\begin{eqnarray}
\frac{M^{\rm phys}}{M^{\rm phys}_{\odot}}&=&
\frac{M}{1.477\times 10^{5}{\rm cm}} \,
\end{eqnarray}
for the physical mass in grams, and
\begin{equation}
f^{\rm phys}=\frac{\omega^{\rm phys}}{2\pi}
 =\frac{c\omega}{2\pi}
=(4.771\times 10^{9}{\rm cm\cdot s^{-1}})\,\omega
\end{equation}
for the frequency in Hertz. Frequencies scale with the central mass proportionally to 
$M^{-1}$; their numerical
values are therefore converted in the physical units by multiplying by the factor
\begin{equation}
 \frac{c}{2\pi M}=(3.231\times 10^{4})\,
 \left(\frac{M}{M_{\odot}}\right)^{\!-1}\,[{\rm{Hz}}].
\end{equation}
For the black hole in Cyg X-1 the current mass estimate \citep{2011ApJ...742...84O} reads
$M^{\rm phys}/M_{\odot}=14.8\pm1$.

From figs. \ref{zobak}--\ref{kridlo} we learn that two peaks occur within only a very narrow parameter range. It is even more 
difficult to produce two peaks of similar height in the $\omega\, S(\omega)$ plot. The condition is that 
two small numbers, $K$ and $1 - \alpha$ have to be equal. For example, in the hard state of Cyg X-1 the two 
most prominent Lorentzians are separated by about two decades in frequency, and they are of identical height, 
which means that in such a state $K \approx 0.01$ and $1 - \alpha = 0.01$. Hence, the two quantities must 
be strongly correlated. Even if the two Lorentzians arise in different regions, the oscillations have to be 
physically linked. This mathematical statement is thus important and has to be taken into account in any 
attempt to explain the nature of the two mechanisms. Naturally, one has to bear in mind that all our 
conclusions are based on a simplified scenario; a more complicated (astrophysically realistic) model
will very likely contain more parameters, which will bring additional degrees of freedom.

We have demonstrated that the power spectra $S_{\rm r}(\omega)$ are double-peaked only for
pairs of $\alpha$ and $K$ within a small area of the parameter
space. However, this does not mean that single-peaked power spectra are lacking any internal 
structure. Unlike the local maximum, it is not obvious how to formalize the presence
of the internal structure into equations.
The simplest continuation of the ideas above is to calculate inflection points of $\omega S(\omega)$,
\begin{equation}
\left.\frac{\rm d^2}{{\rm d}\omega^2}\left[\omega S_{\rm r}(\omega)\right]
\right|_{\omega=\omega_{\rm I1}}=0.
\label{OmegaI1Ror}
\end{equation}
Apparently, equation (\ref{OmegaI1Ror}) has also either one or three positive 
real solutions.

We have been investigating the local maxima of $\omega S(\omega)$, because 
the power spectra $S(\omega)$ of a shot-noise-like processes are peaked at zero and because frequency 
times power versus frequency plots are traditionally used to analyse
of astronomical signals. There is, however, no good a priori reason to prefer the inflection points
of $\omega S(\omega)$ over inflection points of $S(\omega)$ itself,
\begin{equation}
\left.\frac{\rm d^2}{{\rm d}\omega^2}S_{\rm r}(\omega)
\right|_{\omega=\omega_{\rm I0}}=0.
\end{equation}
The regions $C_{{\rm I}0}$ and $C_{{\rm I}1}$ denote areas of the parameter space that exhibit more
than one inflection point of $S_{\rm r}(\omega)$, respectively $\omega S_{\rm r}(\omega)$.
The critical points at their boundaries are connected by blue, and by magenta curve in the figure \ref{zobak}.
These two areas clearly do not coincide, because the multiplication of the power spectra by $\omega$ can in some cases 
remove or create new inflection points. Inspired by the properties of the log-log plotting of power spectra,
we can define the local power-law index of the PSD $q(\omega)$ and associate the internal 
structure of the power spectra with the presence of inflection points,
\begin{eqnarray}
q(\omega)=\frac{{\rm d}\ln(S_{\rm r}(\omega))}{{\rm d}\ln(\omega)}
=\frac{\omega}{S_{\rm r}(\omega)}\frac{{\rm d}S_{\rm r}(\omega)}{{\rm d}\omega},\\
\left.\frac{\rm d^2}{{\rm d}\omega^2}q(\omega)
\right|_{\omega=\omega_{\rm Iq}}=0.
\end{eqnarray}
The part of the parameter space corresponding to power spectra with more than one inflection point
of $q(\omega)$ are within set $C_{{\rm I}q}$ enclosed by the green curve in  
fig. \ref{zobak}, which shows  the positions of the inflection points of the three different types  
for a selected set of $\alpha$s and a whole range of $K$. 
\begin{figure*}[tbh]
\includegraphics[width=0.32\textwidth]{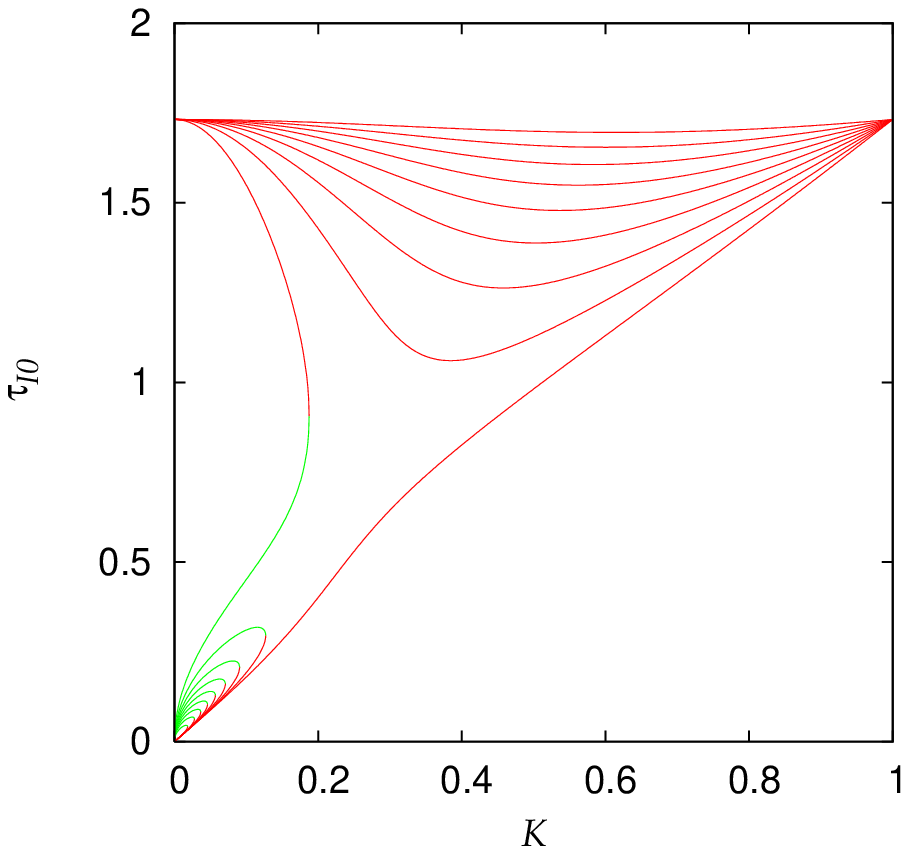}
\includegraphics[width=0.32\textwidth]{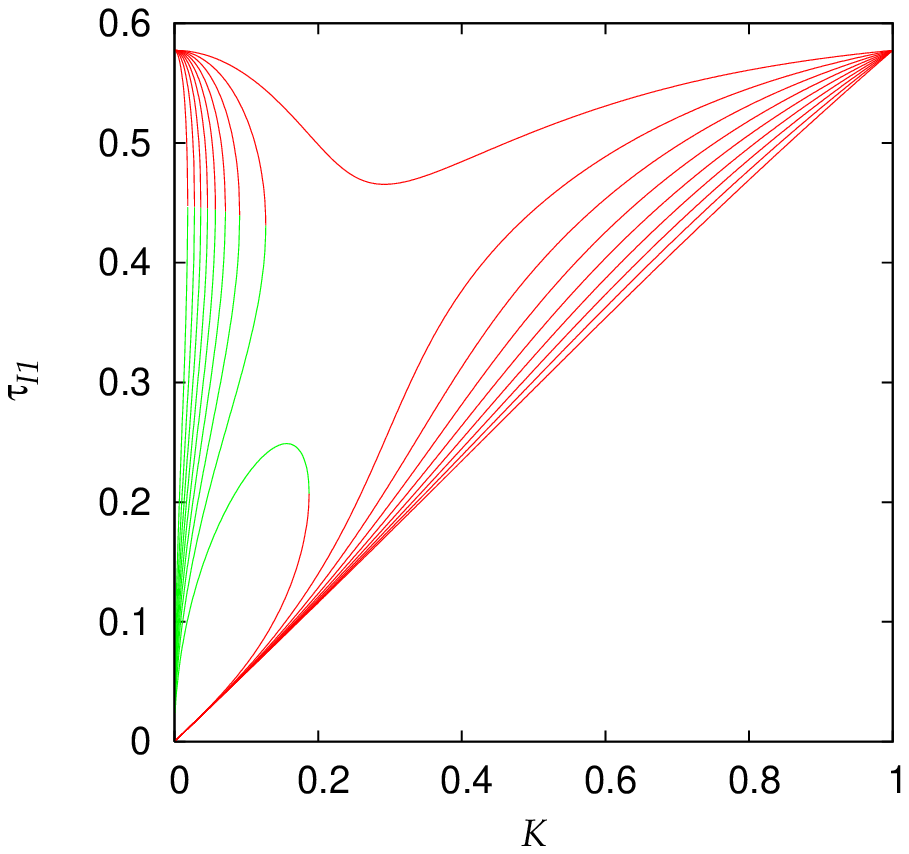}
\includegraphics[width=0.32\textwidth]{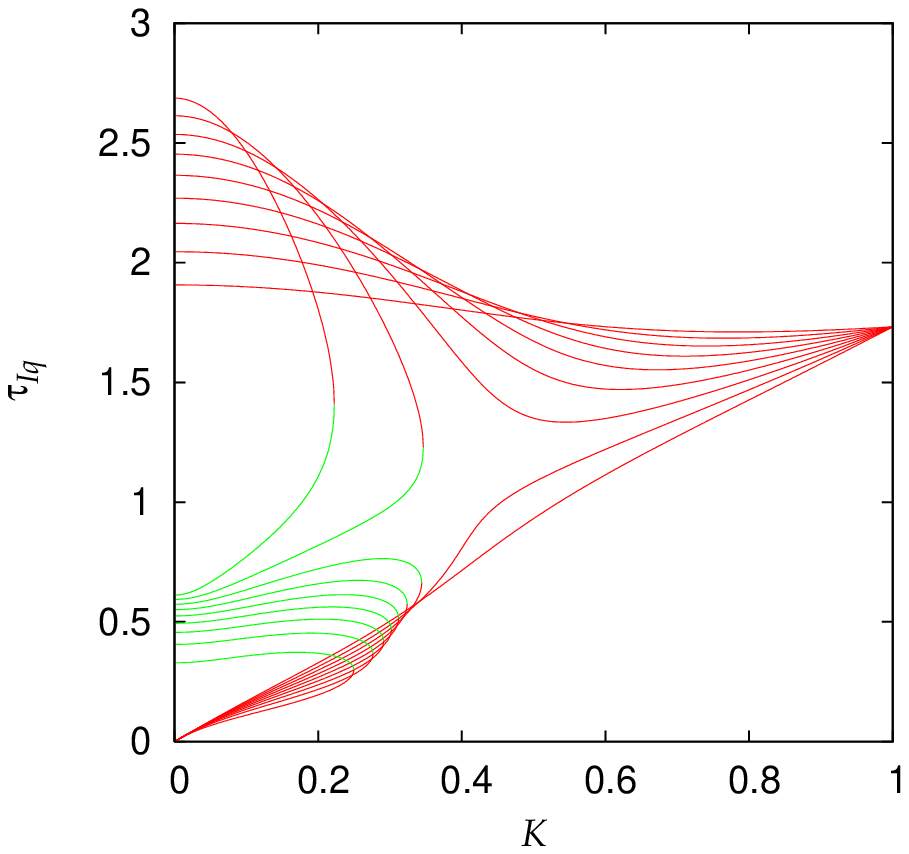}
\caption{
Time-scales $\tau_{\rm I0}$ (left), $\tau_{\rm I1}$ (middle) and $\tau_{\rm Iq}$ (right)
calculated for the PSD given by eq.\ (\ref{2Rorentziany}). We assumed the values of $\alpha$ from 
the set $\{0.1,\,0.2,\,0.3,\,0.4,\,0.5,
\,0.6,\,0.7,\,0.8,\,0.9\}$, and $K$ from the interval $\langle 0,1 \rangle$.}
\label{Inflexy}
\end{figure*}

\subsection{A more general distribution of flare profiles}
We now investigate a more general case with an arbitrary probability distribution $p(\tau)$.
First we observe that the normalization function $I_0(\tau)$ has a similar influence 
on the resulting power spectrum as $p(\tau)$ itself. Therefore, we can without loss of generality
study the power spectrum
\begin{equation}
S(\omega)=\int \frac{1}{1+\tau^2\omega^2} \tilde{p}(\tau){\rm d}\tau,
\label{SimSpec}
\end{equation}
where $\tilde{p}(\tau)=\lambda \tau^2 I_0^2(\tau)p(\tau)$.

\begin{figure*}[tbh]
\includegraphics[width=0.56\textwidth]{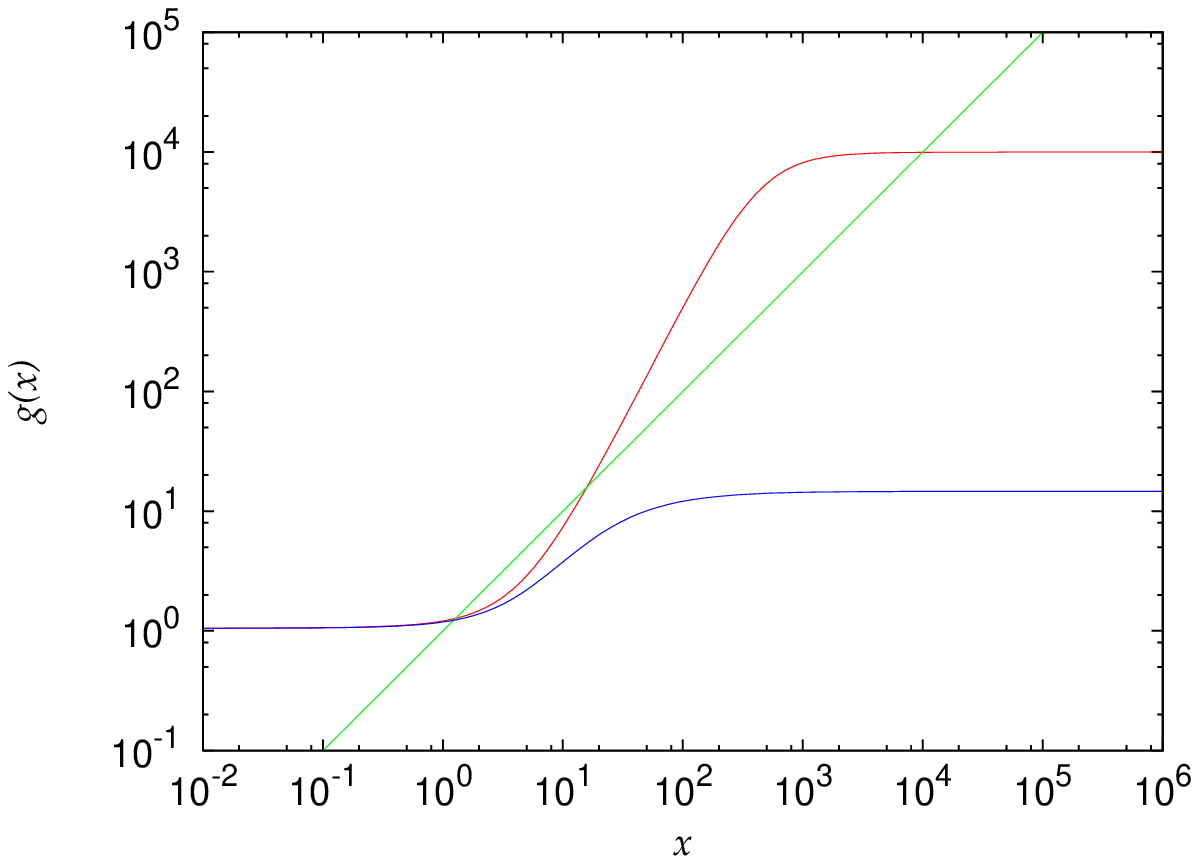}
\includegraphics[width=0.42\textwidth]{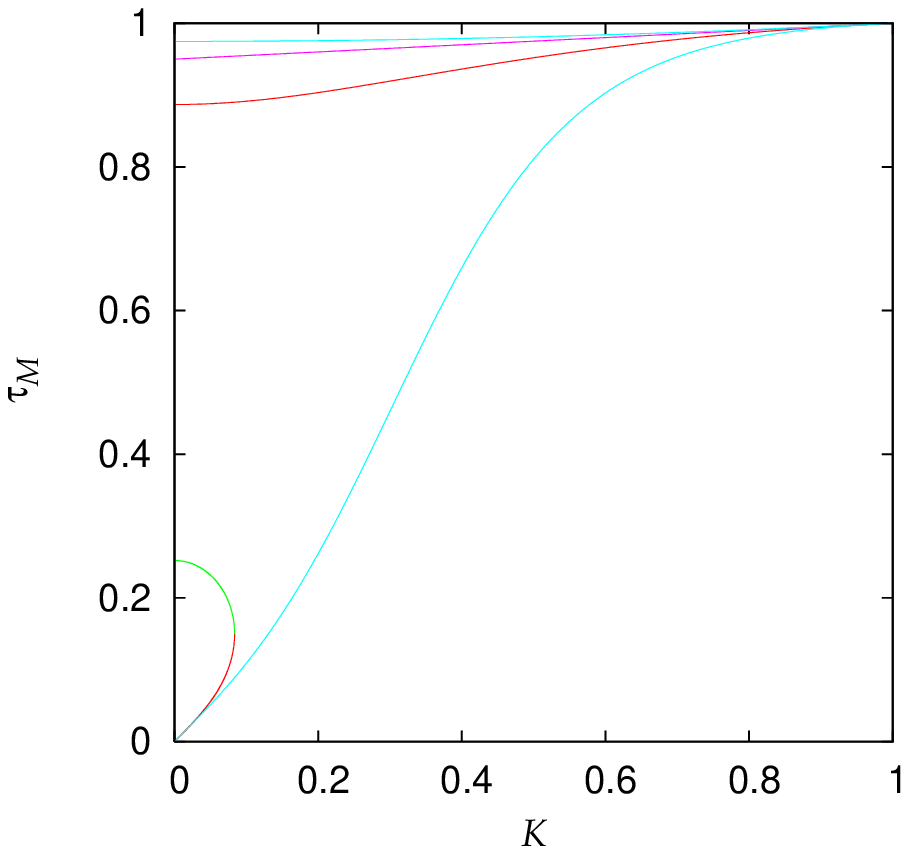}
\caption{
Left: Function $g(x)$ (eq. (\ref{g-def})) for the power spectrum (\ref{2Rorentziany}), calculated
for $\alpha=0.95$, and $K=0.01$ (red) and $K=0.2$ (blue), respectively. The green line represents
the identity function $g(x)=x$.
Right: The corresponding plot of time-scales, similar to the right panel of fig.\ref{kridlo}.
The light-blue lines correspond to the limits from the inequality (\ref{MaxIneq}).}
\label{GaNerovnost}
\end{figure*}

Without any loss of generality we can
set $\lambda=[S(0)]^{-1}$. This rescaling does not change the overall shape of the PSD and 
it normalises $\tilde{p}(\tau)$ to unity. Differentiating $\omega S(\omega)$ from  eq. (\ref{SimSpec}) leads to
\begin{equation}
\frac{\rm d}{{\rm d}\omega}\left[\omega S(\omega)\right]
=\int \frac{1-\tau^2\omega^2}{\left(1+\tau^2\omega^2\right)^2}\, \tilde{p}(\tau){\rm d}\tau
={\rm E}\left[\frac{1-\tau^2\omega^2}{\left(1+\tau^2\omega^2\right)^2}\right].
\label{DiffoSo}
\end{equation}
For the peak frequency we find due to linearity of the averaging operator ${\rm E}[.]$, 
\begin{equation}
{\rm E}\left[\left(1+\tau^2\omega_{\rm M}^2\right)^{-2}\right]
=\omega^2_{\rm M}{\rm E}\left[\tau^2\left(1+\tau^2\omega_{\rm M}^2\right)^{-2}\right].
\end{equation}
Now we substitute $x_{\rm M}$ for $\omega_{\rm M}^2$, define a function $g(x)$ as
\begin{equation}
g(x)=\frac{{\rm E}\left[\left(1+\tau^2x\right)^{-2}\right]}
{{\rm E}\left[\tau^2\left(1+\tau^2x\right)^{-2}\right]},
\label{g-def}
\end{equation}
and observe that the position of the local extreme is determined by the intersection
of $g(x)$ with $x$:
\begin{equation}
g(x_{\rm M})=x_{\rm M}.
\label{FixedPoint}
\end{equation}
It follows from the Jensen inequality that the derivative 
of $g$ is always non-negative (see the appendix). Therefore, the position of $x_{\rm M}$ is constrained 
to the interval $\langle g(0),g(\infty)\rangle$. Straightforwardly, $g(0)={\rm E}\left[\tau^2\right]^{-1}$.
To calculate the upper limit we observe that $1/(1+\tau^2x)\propto \tau^{-2}$ for
$\tau^2 \gg x^{-1}$. Assuming that there is a non-zero lowest time-scale $\tau_{\rm min}>0$ so that $p(\tau)=0$ for $t<\tau_{\rm min}$,
we can replace the Lorentzian in (\ref{g-def}) by $\tau^{-2}$ and put\footnote{We denote $g(\infty)$ as an abbreviation
for a graphically less compact but formally more correct expression, $\lim\limits_{x\rightarrow\infty}g(x)$.}
$g(\infty)={\rm E}\left[\tau^{-4}\right]/{\rm E}\left[\tau^{-2}\right]$.
For the break time-scale we finally find the inequality
\begin{equation}
\sqrt{{\rm E}[\tau^2]}\geq\tau_{\rm M}\geq\sqrt{{\rm E}[\tau^{-2}]/{\rm E}[\tau^{-4}]}.
\label{MaxIneq}
\end{equation}  
Figure \ref{GaNerovnost} illustrates this again in terms of the power spectrum (\ref{2Rorentziany}).
Although $g(x)$ is necessarily a non-decreasing function, 
the intersection $g(x)=x$ needs not to be unique. The inequality (\ref{MaxIneq}) then
holds for all local maxima and minima of the PSD. 

We now further investigate properties of the function $g(x)$. Eq. (\ref{FixedPoint}) states 
that the extrema of the PSD are fixed points of $g(x)$.
We can define the $n$-th iteration of $g(x)$ by recurrent relations,
\begin{equation}
g^{(n)}(x)\equiv g(g^{(n-1)}(x)),\quad
g^{(1)}(x)\equiv g(x).
\end{equation}
It trivially follows from eq. (\ref{FixedPoint}) that the local extrema $x_{\rm M}$ are also fixed points of 
$g^{(n)}(x)$ for all values of $n$. Assuming distributions $\tilde{p}(\tau)$ with 
finite moments $g(0)$ and $g(\infty)$, the function $g(x)$ maps the real half-line $\langle 0,\infty\rangle$
into the finite interval $\langle g(0), g(\infty)\rangle$. From the monotonicity of $g$, it then follows that 
$g$ maps the latter interval within itself.

By repeated applications of $g$ we obtain an infinite sequence of nested inequalities,
\begin{eqnarray}
g(0)\leq g^{(n)}(0)\leq g^{(n+1)}(0),\\
g^{(n+1)}(\infty)\leq g^{(n)}(\infty)\leq g(\infty).
\end{eqnarray}
Because of the Bolzano-Weierstrass theorem, the following limits exist and the same inequality holds
for them as well,
\begin{eqnarray}
g^{(\infty)}(0)=\lim\limits_{n\rightarrow\infty}g^{(n)}(0),\\
g^{(\infty)}(\infty)=\lim\limits_{n\rightarrow\infty}g^{(n)}(\infty),\\
g^{(\infty)}(0)\leq x_{\rm M}\leq g^{(\infty)}(\infty).
\label{GenaralGIneq}
\end{eqnarray}
Both $g^{(\infty)}(0)$ and $g^{(\infty)}(0)$ are fixed points of $g(x)$ and correspond to the lowest and highest local maxima 
of $\omega S(\omega)$, respectively. If they coincide, the power spectrum has a single peak.

The inflection points of the power spectrum can be investigated in a similar way. 
Defining a function
\begin{equation}
h(x)=\frac{{\rm E}\left[\tau^2\left(1+\tau^2x\right)^{-3}\right]}
{{\rm E}\left[\tau^4\left(1+\tau^2x\right)^{-3}\right]},
\label{h-def} 
\end{equation}
it can be easily proven that 
\begin{eqnarray}
\left.\frac{\rm d^2}{{\rm d}\omega^2} S(\omega)
\right|_{\omega=\sqrt{x_{\rm I0}}}=0\,:\,3x_{\rm I0}=h(x_{\rm I0}),\\
\left.\frac{\rm d^2}{{\rm d}\omega^2}\left[\omega S(\omega)\right]
\right|_{\omega=\sqrt{x_{\rm I1}}}=0\,:\,x_{\rm I1}=3h(x_{\rm I1}).
\end{eqnarray}
As we show in the appendix, $h(x)$ is also a non-decreasing function.
Therefore, $h(x)$ has properties similar to $g(x)$, with $h(0)={\rm E}[\tau^2]/{\rm E}[\tau^4]$,
$h(\infty)={\rm E}[\tau^{-4}]/{\rm E}[\tau^{-2}]$ and with a similar structure of
nested subintervals, 
\begin{equation}
\langle h^{(n)}(0),h^{(n)}(\infty)\rangle \supseteq
\langle h^{(n+1)}(0),h^{(n+1)}(\infty)\rangle.
\end{equation}

There are several applications of the previous relations. Clearly, finding the fixed points 
of $g$ and $h$ numerically is as hard as finding the extrema and inflection points of $S(\omega)$ directly.
But by calculating $g(0)$, $h(0)$, $g(\infty)$ and $h(\infty)$, we can roughly estimate the 
position of each feature. Every application of these functions requires calculating two expected
values of some functions of $\tau$. It was demonstrated in the case of two Lorentzians that
these limits are sometimes too wide. In this case we can always determine a tighter 
constraint by applying $g$ and $h$, respectively, on the result at the cost of calculating two
additional expected values per each iteration. Moreover, we can use $g$ and $h$ to find a constraint
on the numbers of local extrema and inflection points.

In the simple case of two Lorentzians we were able to 
calculate directly for which sub-sets of the parameter space of $\alpha$ and $K$ is the 
PSD doubly-peaked and doubly-broken. Let us call these sets  
$C_{\rm M}$, $C_{{\rm I}0}$, $C_{{\rm I}1}$ and $C_{{\rm I}q}$. The parameter space of the general 
case with an arbitrary probability function $\tilde{p}(\tau)$ is infinite-dimensional. In analogy with 
the two-dimensional case of $S_{\rm r}(\omega)$ we can define a subsets of the space of all probability distribution 
functions $\mathcal{C}_{\rm M}$, such that the power spectrum (\ref{SimSpec}) has multiple local extrema $x_{\rm M}$  
if and only if $\tilde{p}(\tau)\in \mathcal{C}_{\rm M}$. 

Testing if a particular function $\tilde{p}(\tau)$ is contained in $\mathcal{C}_{\rm M}$ means 
calculating the power spectrum directly. However, using the function $g(x)$, we can construct 
approximation sets $\mathcal{C}_{\rm M}^{\rm N}$ and 
$\mathcal{C}_{\rm M}^{\rm S}$ representing the necessary and sufficient conditions on $\tilde{p}(\tau)$
to produce a multiply peaked power spectrum, $\mathcal{C}_{\rm M}^{\rm S}\subseteq\mathcal{C}_{\rm M}
\subseteq\mathcal{C}_{\rm M}^{\rm N}$. 
Using $g$ and $h$, the approximating sets can be defined by an inequality, or a set of inequalities 
between some moments of $\tau$. We now present two examples of such constructions and apply 
them to the example of the PSD (\ref{2Rorentziany}).

\subsection{Examples of the construction procedure}
\begin{figure*}[tbh]
\includegraphics[width=0.48\textwidth]{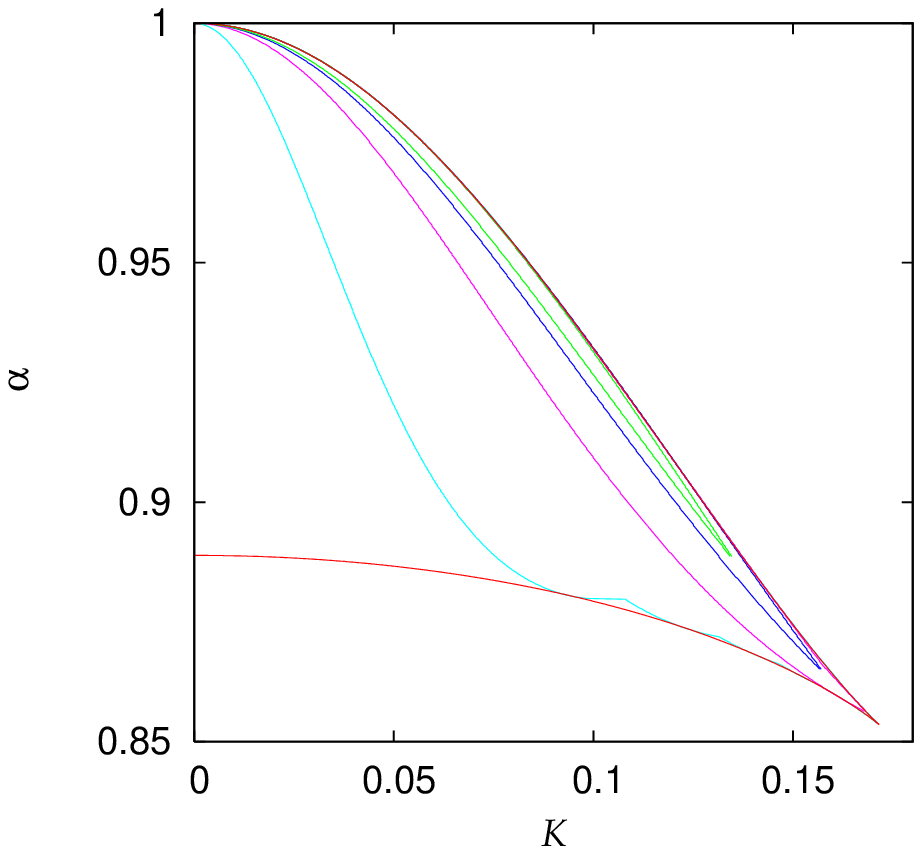}
\hfill
\includegraphics[width=0.45\textwidth]{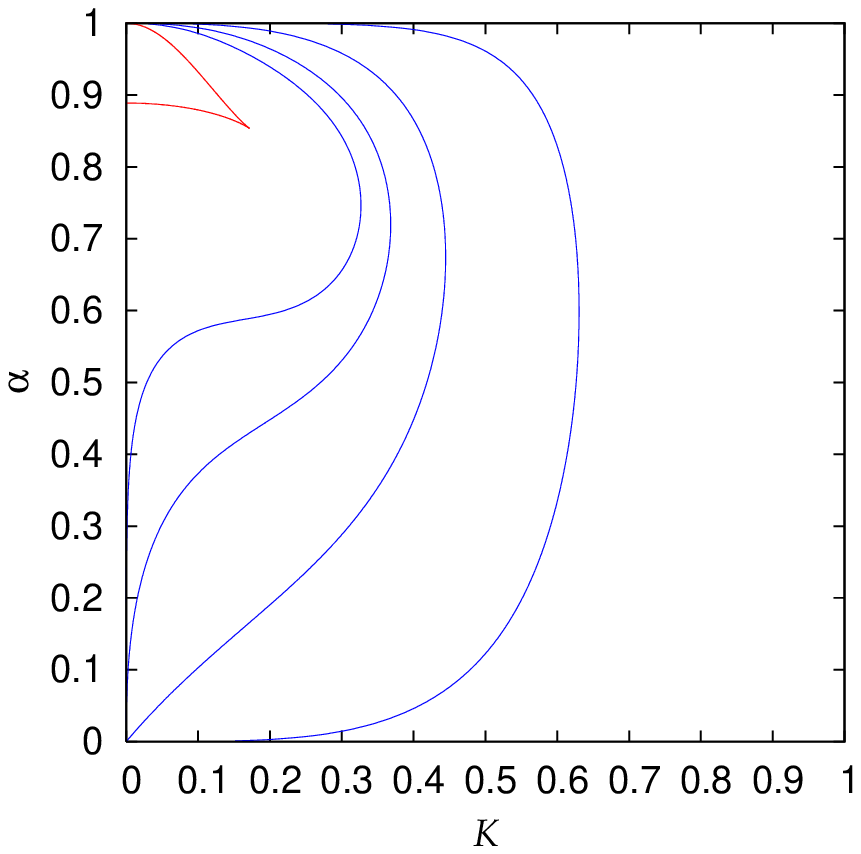}\\[25pt]
\includegraphics[width=0.49\textwidth]{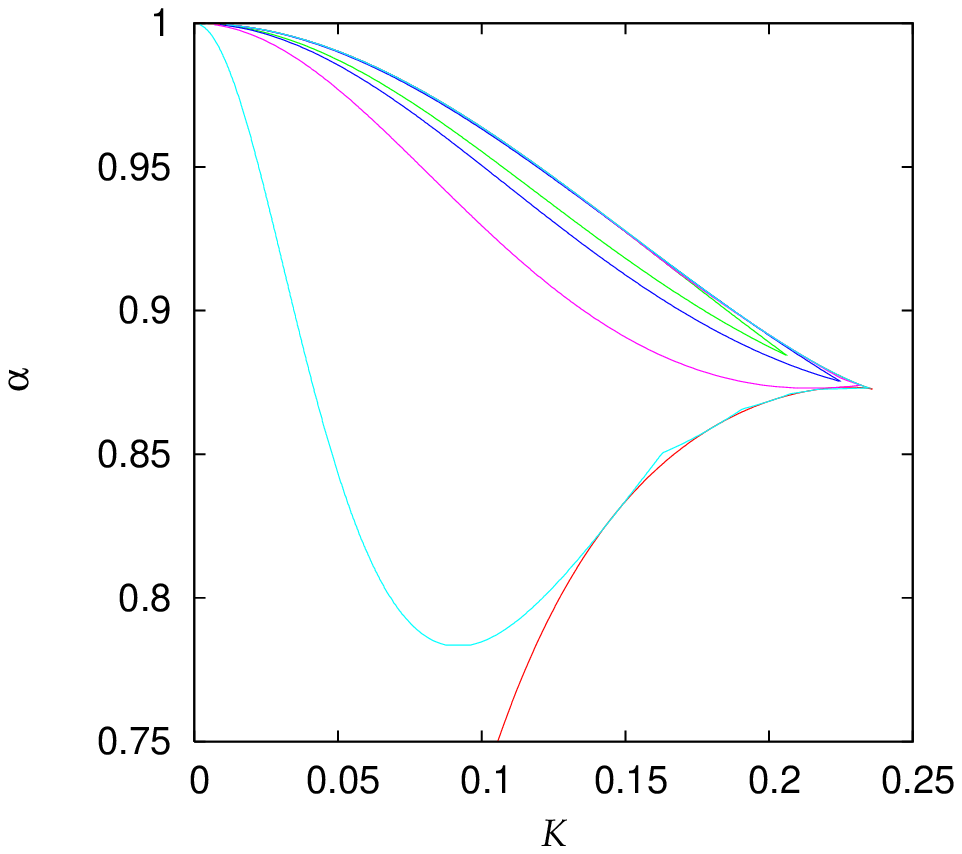}
\hfill
\includegraphics[width=0.49\textwidth]{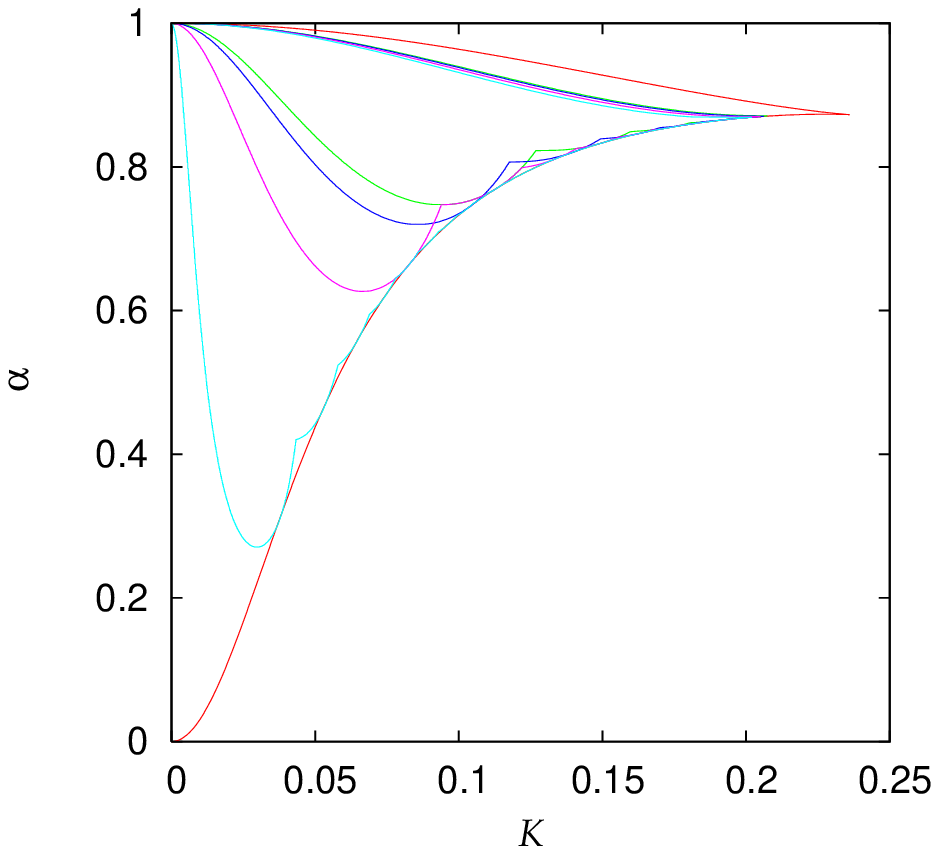}
\caption{
Top left: Boundaries of approximating sets $\mathcal{C}_{\rm M}^{\rm S}$ for the power spectrum $S_{\rm r}(\omega)$. 
The function $g(x)$ was evaluated in $L$ points, evenly placed in between $g(0)$ and $g(\infty)$. The nested curves 
correspond to $\mathcal{C}_{\rm M}^{\rm S}$ calculated for $L$ equal to $4$, $5$, $9$, and $49$. Top right: The nested
sets $\mathcal{C}_{\rm M}^{\rm N}$ calculated from the upper limit of $g'(x)$. We have taken $x_+=g^{(n)}(0)$ and
$x_-=g^{(n)}(\infty)$ for $n$ from $1$ to $4$. Bottom left: The same as in the top left panel for sets
$\mathcal{C}_{\rm I1}^{\rm S}$. Bottom right: The same as in the bottom left panel, this time with a
grid of points that samples the area around point $3h(0)$ more densely but that does not cover the whole interval
$\langle 3h(0),3h(\infty)\rangle$. The values of $L$ are the same as in the previous examples. The approximating sets
$\mathcal{C}_{\rm I1}^{\rm S}$ cover a broader part of the whole set $\mathcal{C}_{\rm I1}$. However, they omit
a small area of high $\alpha$. This demonstrates that an adaptive grid can perform significantly better than a regular
one.}
\label{ApproxOfCs}
\end{figure*}

We take $x_+$ and $x_-$ such that $x_+<x_{\rm M}<x_-$ for all fixed points 
$g(x_{\rm M})=x_{\rm M}$. By construction,  $g(x_+)>x_+$ and $g(x_-)<x_-$.  
If there is more than one fixed point $x_{\rm M}$, the function $s_g(x)=g(x)-x$ changes its sign in the interval 
$\langle x_+, x_-\rangle$  more than once. We can pick $L$ values $x_j$ from the interval
such that $x_+<x_1<\dots<x_L< x_-$. If the sequence of $s_g(x_j)$ changes sign more than once,
the fixed point is not unique. Figure \ref{ApproxOfCs} shows the application of the test on the power spectrum  
(\ref{2Rorentziany}) for $x_+=g(0)$, $x_-=g(\infty)$ and $x_j$ placed uniformly over the whole
interval. The results are plotted for $L$ equal to $4$, $5$, $9$ and $50$. Note that this is not 
the only possible choice of the $x$. All $x_+=g^{(n)}(0)$ and $x_-=g^{(n)}(\infty)$
are admissible, and likewise, $x_j$ can be chosen in some irregular or even adaptive way by using first $L$
steps of some numerical root-finding method (e.g. the regula falsi method). What matters is the final complexity 
of the test. It produces a set of $L$ inequalities that must be satisfied by $\tilde{p}(\tau)$ to be in
$\mathcal{C}_{\rm M}^{\rm S}$.

For set $\mathcal{C}_{\rm M}^{\rm N}$ we can use a different approach. The well-known Banach fixed-point
theorem says that if the function $g$ is a contracting mapping, i.e. if there exists a number $0\leq\lambda<1$
such that $|g(x)-g(y)|<\lambda|x-y|$ for all admissible $x$ and $y$, then the fixed-point $x_{\rm M}$ is unique.
Because $g(x)$ is a continuous and non-decreasing function, it follows from the mean value theorem that it is 
a contracting mapping if the maximum of its derivative is lower than one, $g'(x)<1$. Defining a new function $\ell(x)$ as
\begin{equation}
\ell(x)=\frac{{\rm E}\left[(1+\tau^2 x)^{-3}\right]}{{\rm E}\left[\tau^4(1+\tau^2 x)^{-3}\right]},
\end{equation}
we can write the derivative of $g$,
\begin{equation}
g'(x)=2\frac{\ell(x)-h^2(x)}{\left(h(x)+x\right)^2}.
\label{gDerivByL}
\end{equation}
Similarly to $g$ and $h$, also $\ell(x)$ is a non-decreasing function that can be expressed in terms of the functions
$g$ and $h$,
\begin{equation}
\ell(x)=g(x)\left(h(x)+x\right)-xh(x).
\label{EllDeff}
\end{equation}
Because the function $g$ maps the whole semi-axis $\langle 0,\infty\rangle$ into $\langle g(0),g(\infty)\rangle$ it is sufficient
to investigate the derivative $g'(x)$ only on the latter interval. Because $g$, $h$, and $\ell$ are non-decreasing  
functions, the following inequality holds for all $x_+$, $x_-$ and $y$ that satisfy $g(0)\leq x_+\leq y\leq x_-\leq g(\infty)$,
\begin{equation}
g'(y)\leq D_{\rm max}(x_+,x_-)=2\frac{l(x_-)-h^2(x_+)}{\left(h(x_+)+x_+\right)^2}.
\end{equation}
Using this inequality, we can take for $\mathcal{C}^{\rm N}_{\rm M}$ the set of all distributions $\tilde{p}(\tau)$
for which $D_{\rm max}(x_+,x_-)> 1$. Figure \ref{ApproxOfCs}  demonstrates this on the bi-Lorentzian power spectrum. 
We employed $x_+=g^{(n)}(0)$ and $x_-=g^{(n)}(\infty)$ with $n$ from $1$ to $4$. The margins
derived from this criterion can be relatively wide, both because $D_{\rm max}(x_+,x_-)$ overestimates
the maximum of $g'(x)$, and also because some functions $g(x)$ can have a single fixed point $x_{\rm M}$ and a 
derivative $g'(y)>1$ ($y\not=x_{\rm M}$) at the same time. On the other hand, if $D_{\rm max}(x_+,x_-)<1$, 
$g(x)$ surely exhibits a single fixed point.

Unfortunately, the analysis of inflection points of the local power-law slope cannot be generalized
in the same way. However, we note that $q(\omega)$ is related to $g$ by
\begin{equation}
q(\omega)=\frac{-2\omega^2}{\omega^2+g(\omega^2)}.
\label{qDeff}
\end{equation}
From here we can immediately find that $q(0)=0$ and $q(\infty)=-2$. Furthermore, from (\ref{FixedPoint}) we see that $q(\omega_{\rm M})=-1$ 
and conversely, by expressing $g$ in terms of $q$, we can see that all $\omega$ with $q(\omega)=-1$ are the
fixed points of $g$. This is not surprising since in the vicinity of the local extreme 
$(\omega_{\rm M}+y) S(\omega_{\rm M}+y)={\rm const.}+o(y)$. It is, however, a good motivation 
to briefly investigate the behaviour of the function $\omega^\gamma S(\omega)$ in terms of its local extrema 
$\omega_{{\rm M}\gamma}$ and inflection-points $\omega_{{\rm I}\gamma}$,
\begin{equation}
\left.\frac{\rm d}{{\rm d}\omega} \left[\omega^\gamma S(\omega)\right]\right|_{\omega=\sqrt{x_{{\rm M}\gamma}}}=0,\quad
\left.\frac{\rm d^2}{{\rm d}\omega^2} \left[\omega^\gamma S(\omega)\right]\right|_{\omega=\sqrt{x_{{\rm I}\gamma}}}=0.
\end{equation}
Following the same procedure that led to the equations for $x_{\rm M}$, $x_{\rm I0}$, and $x_{\rm I0}$, we obtain
\begin{eqnarray}
\gamma g(x_{{\rm M}\gamma})-(2-\gamma)x_{{\rm M}\gamma}=0,\\
A_\gamma x_{{\rm I}\gamma}^2+B_\gamma h(x_{{\rm I}\gamma})x_{{\rm I}\gamma}+C_\gamma l(x_{{\rm I}\gamma})=0,
\label{HGamma}
\end{eqnarray}
where $A_\gamma=(\gamma-2)(\gamma -3)$, $B_\gamma=2(\gamma^2-3\gamma-1)$ and $C_\gamma=\gamma(\gamma-1)$.
We see that the solution of the problem is fully determined by the functions $g$ and $h$. 

This generalization brings nothing new for the local extrema $x_{{\rm M}\gamma}$. The constraints for the number 
of fixed points $x_{{\rm M}\gamma}$ and methods for approximating their positions can be obtained from 
the corresponding methods for $\gamma=1$ by replacing $g(x)$ with $g_\gamma(x)=\gamma(2-\gamma)^{-1}g(x)$ 
and the local slope at the extremal points is $q(\omega_{{\rm M}\gamma})=-\gamma$, 
as expected. However, the problem of the inflection points is more complicated with the exceptions of $\gamma=0$ and $\gamma=1$. 
A possible generalization of $h(x)$ for $\gamma\in\langle0,1\rangle$ is given by
\begin{equation}
h_\gamma(x)=-\frac{B_\gamma}{2A_\gamma}h(x)+\sqrt{\frac{B^2_\gamma}{4A^2_\gamma}h^2(x)-\frac{C_\gamma}{A_\gamma}\ell(x)}.
\label{HGammaSolv}
\end{equation}
\section{Polynomial flares}
In all of the previous considerations, we have assumed that the flare profiles are given by decaying exponentials 
and the distribution $\tilde{p}(\tau)$ is positive and can be normalized
to one. No other assumptions on $\tilde{p}(\tau)$ were imposed. Therefore, the resulting inequalities
are very general. We now approach the case of flares of the form
\begin{equation}
I_k(t|\tau)=I_0(\tau)P_k\left(\frac{t}{\tau}\right)\exp\left(-\frac{t}{\tau}\right)\theta(t),
\end{equation}
where $P_k$ is a polynomial of $k$-th order, and $I_0(\tau)$ is an arbitrary non-negative and quadratically
integrable function. As a motivation we took $I_0(\tau)=1/\tau$, $k=1$ and $P_1(t/\tau)=t/\tau$ and calculated 
a power spectrum of a mixture of these profiles with the probability distribution $p(\tau)$,
\begin{equation}
S(\omega)=\int\limits_{\tau_{\rm min}}^{\tau_{\rm max}}S(\omega|\tau)p(\tau){\rm d}\tau=
\int\limits_{\tau_{\rm min}}^{\tau_{\rm max}}\frac{p(\tau)}{\left(1+\tau^2\omega^2\right)^2}{\rm d}\tau.
\label{TExpSp}
\end{equation} 
It can be easily proven that the term $S(\omega|\tau)$ from the previous equation satisfies
a relation  
\begin{equation}
S(\omega|\tau)=\left(1+\tau^2\omega^2\right)^{-2}=\frac{1}{2\tau}\frac{{\rm d}}{{\rm d}\tau}\tau^2
\left(1+\tau^2\omega^2\right)^{-1}.
\end{equation}
Using this relation, we can rewrite equation (\ref{TExpSp}) by integration by parts,
\begin{eqnarray}
S(\omega)&=&\left[\frac{\tau}{2}p(\tau)\left(1+\tau^2\omega^2\right)^{-1}\right]_{\tau_{\rm min}}^{\tau_{\rm max}}
\nonumber\\&+&\frac{1}{2}\int\limits_{\tau_{\rm min}}^{\tau_{\rm max}}\frac{p(\tau)-\tau\frac{{\rm d}}{{\rm d}\tau} p(\tau)}
{1+\tau^2\omega^2}{\rm d}\tau.
\label{PerPartes1}
\end{eqnarray}
We observe that the power spectrum can be rewritten in the form
\begin{eqnarray}
S(\omega)&=&\int\limits_{\tau_{\rm min}}^{\tau_{\rm max}}\frac{p_2(\tau)}
{1+\tau^2\omega^2}{\rm d}\tau,\\
p_2(\tau)&=&\frac{1}{2}\Big[\tau p(\tau)\left(\delta(\tau-\tau_{\rm max})
-\delta(\tau-\tau_{\rm min})\right)\nonumber\\  &+&p(\tau)-\tau\frac{{\rm d}}{{\rm d}\tau} p(\tau)\Big].
\end{eqnarray}
If $p(\tau_{\rm min})=0$ and $p(\tau)-\tau p'(\tau)\geq 0$ for all $\tau\in\langle\tau_{\rm min},\tau_{\rm max}\rangle$,
$p_2(\tau)$ is behaving well, can be normalized to unity, and the power spectrum can be studied in terms of 
the functions $g$ and $h$, as described previously, with the only difference that all averages 
must be calculated using $p_2(\tau)$ instead of $p(\tau)$. 

It is not apparent on first sight how to interpret these average values. Therefore, 
we now study the general case of an arbitrary polynomial
(\ref{TExpSp}) assuming that the power spectrum can be reduced to the pure-exponential case. The PSD of 
a single profile $(\ref{TExpSp})$ can be written as
\begin{equation}
S(\omega|\tau)=\left|\hat{I}_k(\omega|\tau)\right|^2=\frac{Q_k(\tau^2\omega^2)}{\left(1+\tau^2\omega^2\right)^{k+1}},
\end{equation}
where $Q_k(z)$ is a $k$-th order polynomial, different than (but related to) $P_k(t)$. We define
 $\mathbf{\hat{Q}}_k$ by
\begin{equation}
\mathbf{\hat{Q}}_k\frac{1}{1+\tau^2\omega^2}=\frac{Q_k(\tau^2\omega^2)}{\left(1+\tau^2\omega^2\right)^{k+1}}.
\end{equation}
The operator is constructed entirely from $\tau$, derivatives over $\tau$, and from the coefficients of $P_k$
(see the appendix). 

The generalization of equation (\ref{PerPartes1}) is
\begin{eqnarray}
S(\omega)&=&\int\limits_{\tau_{\rm min}}^{\tau_{\rm max}}\tilde{p}(\tau)
\frac{Q_k(\tau^2\omega^2)}{\left(1+\tau^2\omega^2\right)^{k+1}}{\rm d}\tau
=\int\limits_{\tau_{\rm min}}^{\tau_{\rm max}}
\tilde{p}(\tau)\mathbf{\hat{Q}}_k\frac{1}{1+\tau^2\omega^2}{\rm d}\tau\nonumber\\
&=&\int\limits_{\tau_{\rm min}}^{\tau_{\rm max}}
\frac{1}{1+\tau^2\omega^2}\mathbf{\hat{Q}}^\dagger_k \tilde{p}(\tau) {\rm d}\tau
=\int\limits_{\tau_{\rm min}}^{\tau_{\rm max}}
\frac{p_Q(\tau)}{1+\tau^2\omega^2}{\rm d}\tau,
\end{eqnarray}
where $\mathbf{\hat{Q}}^\dagger_k$ is the Hermitian conjugate of $\mathbf{\hat{Q}}_k$.
For $p_Q(\tau)$ positive over the interval of all admissible $\tau$s, we can calculate the function 
$g(x)$,
\begin{equation}
g(x)=\frac{{\rm E}_Q\left[\left(1+\tau^2x\right)^{-2}\right]}
{{\rm E}_Q\left[\tau^2\left(1+\tau^2x\right)^{-2}\right]}=
\frac{{\rm E}\left[\mathbf{\hat{Q}}_k\left(1+\tau^2x\right)^{-2}\right]}
{{\rm E}\left[\mathbf{\hat{Q}}_k\tau^2\left(1+\tau^2x\right)^{-2}\right]},
\label{gQ}
\end{equation}
where ${\rm E}_Q\left[.\right]$ is an averaging operator with respect to the distribution $p_Q(\tau)$.
We note that we do not have to compute the function $p_Q(\tau)$. 

As $\mathbf{\hat{Q}}_k$ operates with $\tau$ only, it also commutes with the derivative in $\omega$. Therefore,
we can easily calculate the numerator and denominator of (\ref{gQ}),
\begin{eqnarray}
\mathbf{\hat{Q}}_k\frac{\tau^2}{\left(1+\tau^2\omega^2\right)^{2}}&=&-\frac{1}{2\omega}\frac{{\rm d}}{{\rm d}\omega}S(\omega|\tau),\\
\mathbf{\hat{Q}}_k\frac{1}{\left(1+\tau^2\omega^2\right)^{2}}&=&\frac{{\rm d}}{{\rm d}\omega}\omega S(\omega)
-\frac{\omega}{2}\frac{{\rm d}}{{\rm d}\omega}S(\omega|\tau). 
\label{gQcit}
\end{eqnarray}
The value of $g(0)$ is related to the mean moment profiles as
\begin{eqnarray}
g(0)&=&\frac{{\rm E}\left[E_I^2(\tau)\right]}{{\rm E}\left[E_I^2(\tau)W^2_I(\tau)\right]},\\
E_I(\tau)&=&\int\limits_0^\infty I(t|\tau){\rm d}t,\\
W^2_I(\tau)&=&E_I^{-1}(\tau)\int\limits_0^\infty t^2I(t|\tau){\rm d}t
-\left[E_I^{-1}(\tau)\int\limits_0^\infty tI(t|\tau){\rm d}t\right]^2.
\label{WidthI}
\end{eqnarray} 
The function $E_I(\tau)$ can be interpreted as the total energy emitted by the flare and $W_I(\tau)$
is a measure of the width of the flare in the temporal domain. Equation (\ref{WidthI}) is formally
identical with the prescription for the standard deviation.
For the high-frequency limit we find
\begin{equation}
g(\infty)=\frac{\left((k+1)q_k-q_{k-1}\right){\rm E}\left[\tau^{-4}\right]}{q_k{\rm E}\left[\tau^{-2}\right]},
\end{equation}
where $q_l$ are the coefficients of the polynomial $Q_k(z)=\sum_{l=0}^k q_l z^l$. 
The situation of $h(x)$ is completely analogical. We can use the commutation properties of $\mathbf{\hat{Q}}_k$
and express $h(\omega^2)$ as
\begin{equation}
h(\omega^2)=\frac{{\rm E}\left[S''(\omega|\tau)-\frac{3}{\omega}[\omega S(\omega|\tau)]''\right]}
{{\rm E}\left[\frac{3}{\omega^2}S''(\omega|\tau)-\frac{1}{\omega^2}[\omega S(\omega|\tau)]''\right]}.
\end{equation}

The asymptotic slope of the power spectra $S(\omega |\tau)$ is $q(\infty)=-2(l+1)$, where $l$ is the 
order of the lowest non-zero coefficient of the polynomial $P_k(t)$. Apparently, such a case can not 
be transformed onto the purely exponential case, because its asymptotical slope (\ref{qDeff}) is always $q(\infty)=-2$.
However, we can easily repeat the whole analysis for flares of the form $I(t,\tau)=(t^k/\tau^{k+1}) \exp(-t/\tau)\,
\theta(t)$, leading to the power spectrum 
\begin{equation}
S_k(\omega)={\rm E}\left[\frac{(k!)^2}{\left(1+\tau^2\omega^2\right)^{k+1}}\right].
\label{genSpecTk}
\end{equation}
For every $k$ we can define the functions $g_k$ and $h_k$,
\begin{eqnarray}
g_k(x)&=&\frac{{\rm E}\left[\left(1+\tau^2x\right)^{-k-2}\right]}
{{\rm E}\left[\tau^2\left(1+\tau^2x\right)^{-k-2}\right]},\\
h_k(x)&=&\frac{{\rm E}\left[\tau^2\left(1+\tau^2x\right)^{-k-3}\right]}
{{\rm E}\left[\tau^4\left(1+\tau^2x\right)^{-k-3}\right]}.
\end{eqnarray}
It is easy to prove by direct calculation that the extrema and inflection points of $\omega S_k(\omega)$
and $S_k(\omega)$ satisfy the relations
\begin{eqnarray}
x_{\rm M}&=&\frac{1}{2k+1}g_k(x_{\rm M}),\\ 
x_{\rm I0}&=&\frac{1}{2k+3}h_k(x_{\rm I0}),
\quad x_{\rm I1}=\frac{3}{2k+1}h_k(x_{\rm I1}).
\end{eqnarray}
Both $g_k$ and $h_k$ are non-decreasing (see the appendix) and can therefore be used 
analogously to $g$ and $h$. Apparently, $g_k(0)=g(0)$ and $h_k(0)=h(0)$,
for the upper limits we find,
\begin{equation}
g_k(\infty)=h_k(\infty)=\frac{{\rm E}\left[\tau^{-4-2k}\right]}
{{\rm E}\left[\tau^{-2-2k}\right]}.
\end{equation}
Finally, the local slope of $S_k(\omega)$ is related to $g_k(x)$ by the formula
\begin{equation}
q_k(\omega)=\frac{-(2k+2)\omega^2}{g_k(\omega)+\omega^2}.
\end{equation}

\subsection{Alternative forms of the function $g$} 
In the previous sections we described the properties of the function $g$. 
By construction, $g$-values are in every point directly related to the 
power spectrum and its derivatives; see equations (\ref{qDeff}) and (\ref{gQ})--(\ref{gQcit}).
However, in all previous applications we have required only that $g$ is a non-decreasing function,
$g(0)$ and $g(\infty)$ are positive finite numbers, and its fixed points $g(x_{\rm M})=x_{\rm M}$ 
correspond to the local extrema of $\omega S(\omega)$. Clearly, for given $S(\omega)$, these 
conditions do not fix the function uniquely. For instance, every member of the sequence 
$g^{(n)}(x)$ has the required properties. It is therefore natural to ask whether there is an alternative, 
easily calculable form of $g$. 

A simple set of solutions to this problem can be found by adding a zero to 
equation (\ref{DiffoSo}) to obtain
\begin{equation}
\frac{\rm d}{{\rm d}\omega}\left[\omega S(\omega)\right]
={\rm E}\left[\frac{1-(\tau^2+f(\omega^2|\tau)-f(\omega^2|\tau))\omega^2}{\left(1+\tau^2\omega^2\right)^2}\right]=0,
\end{equation}
where $f(\omega^2|\tau)$ is a function of $\omega^2$ and $\tau$. Properties of  $f(\omega^2|\tau)$
are specified below. Due to linearity of the averaging operator we can define 
a new function $g_{_{[f]}}(x)$ as
\begin{equation}
g_{_{[f]}}(x)=\frac{{\rm E}\left[(1+xf(x|\tau))\left(1+\tau^2x\right)^{-2}\right]}
{{\rm E}\left[(\tau^2+f(x|\tau))\left(1+\tau^2x\right)^{-2}\right]}.
\label{gf-def}
\end{equation}
By construction, the functions $g_{_{[f]}}(x)$ and $g(x)$ have identical fixed points irrespective of
the choice of $f(x|\tau)$. The limits of $g_{_{[f]}}$ are given by
\begin{eqnarray}
g_{_{[f]}}(0)={\rm E}\left[\tau^2+f(0|\tau)\right]^{-1},\\
g_{_{[f]}}(\infty)=\frac{{\rm E}\left[\tau^{-4}(1+\phi(\tau))\right]}{{\rm E}\left[\tau^{-2}\right]},
\end{eqnarray}
where $\phi(\tau)$ is the limit
\begin{equation}
\phi(\tau)=\lim\limits_{x\rightarrow\infty} xf(x|\tau).
\end{equation}
Both $f(0|\tau)$ and $\phi(\tau)$ must be finite and chosen such that
$0<g_{_{[f]}}(0)\leq g_{_{[f]}}(\infty)<\infty$. Unlike the previous cases, the positivity
of the derivative of $g_{_{[f]}}$ does not follow from any theorem and must be
checked separately for every particular choice of $f$.

\section{Exponential flares with periodic modulation}
\begin{figure*}[tbh]
\includegraphics[width=0.49\textwidth]{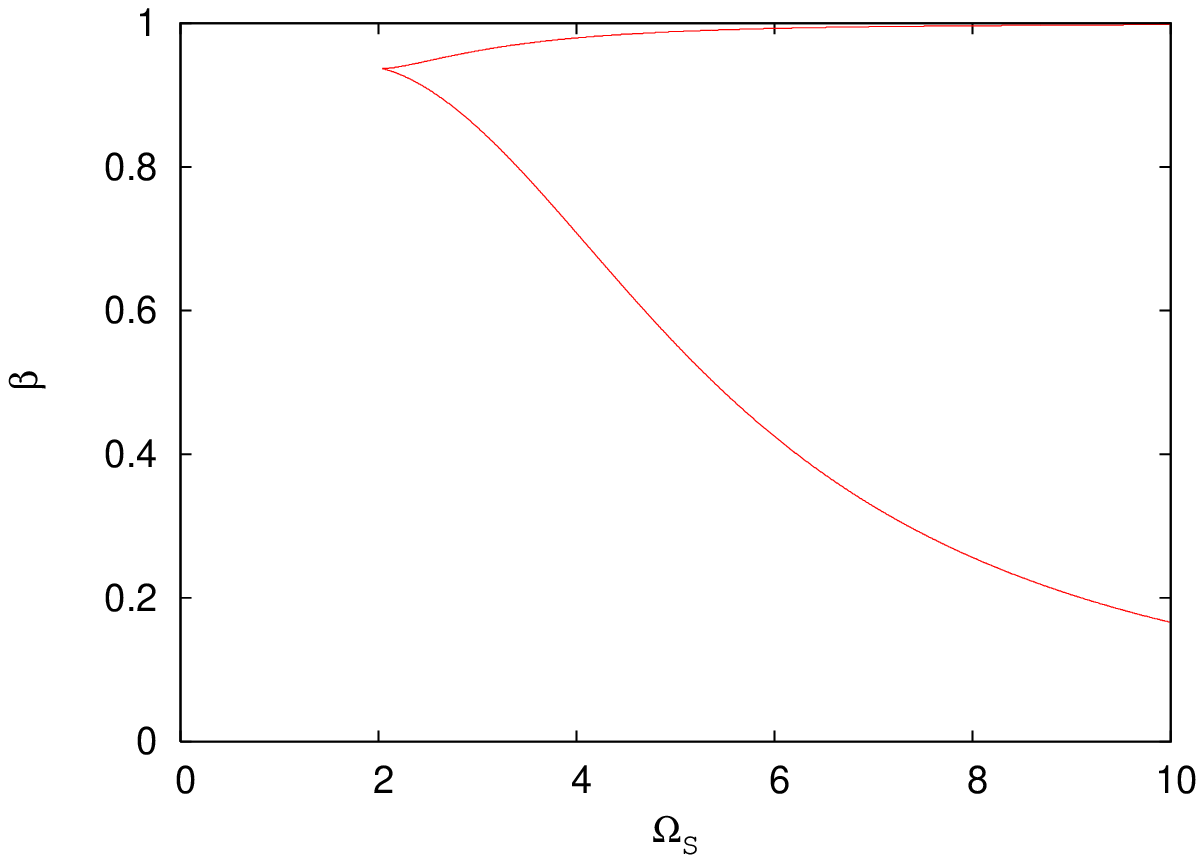}
\includegraphics[width=0.49\textwidth]{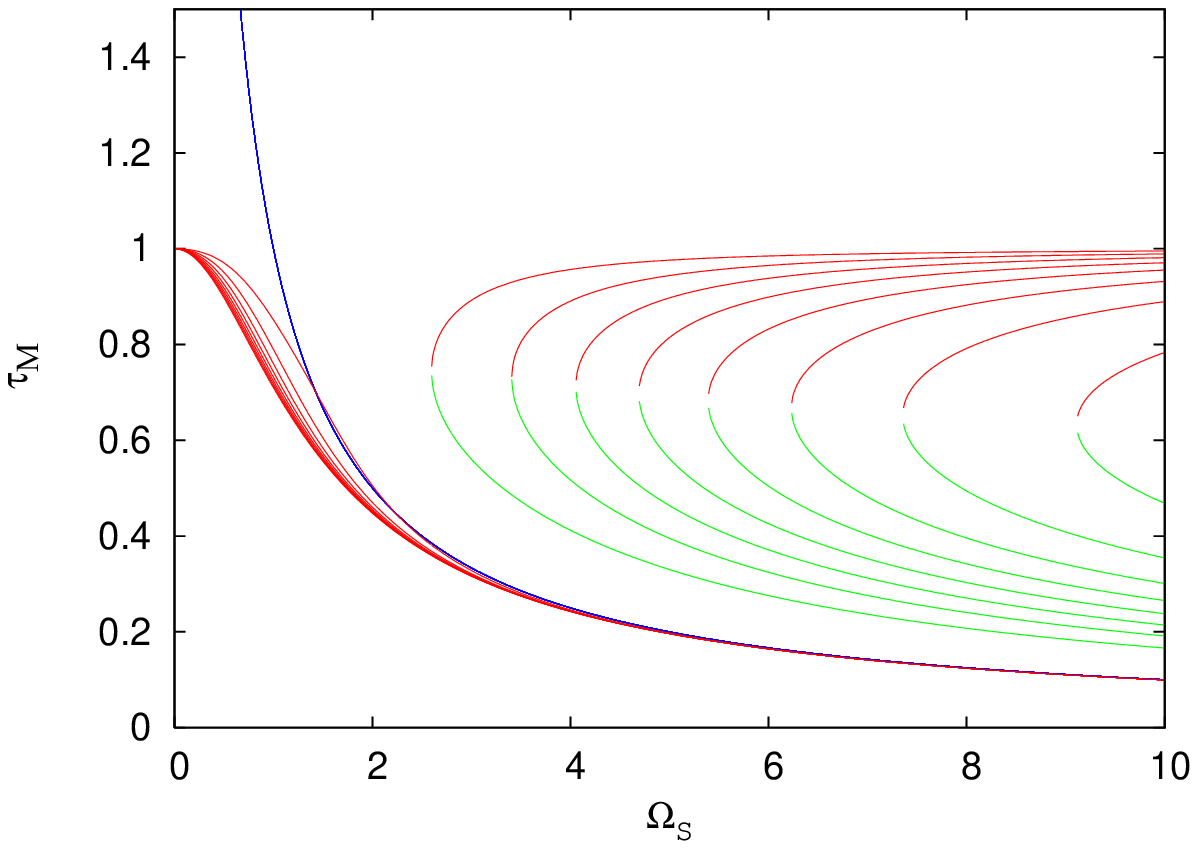}
\caption{Left: 
Boundary of set $C_{\rm M}$ of doubly-peaked power spectra $\omega S_{\rm rD}(\omega)$.
Right: The time-scales $\tau_{\rm M}$ calculated for $\beta$ from the set $\{0.1,\,0.2,\,0.3,\,0.4,\,0.5,
\,0.6,\,0.7,\,0.8,\,0.9\}$ and for $\Omega_{\rm s}$ from the interval $\langle 0, 10\rangle$.
The blue line corresponds to $\Omega_{\rm s}^{-1}$. Again, the red lines denote the $\tau_{\rm M}$
of the peaks, and the green lines denote the local minimum.}
\label{MaxLorDopp}
\end{figure*}

\begin{figure*}[tbh]
\includegraphics[width=0.32\textwidth]{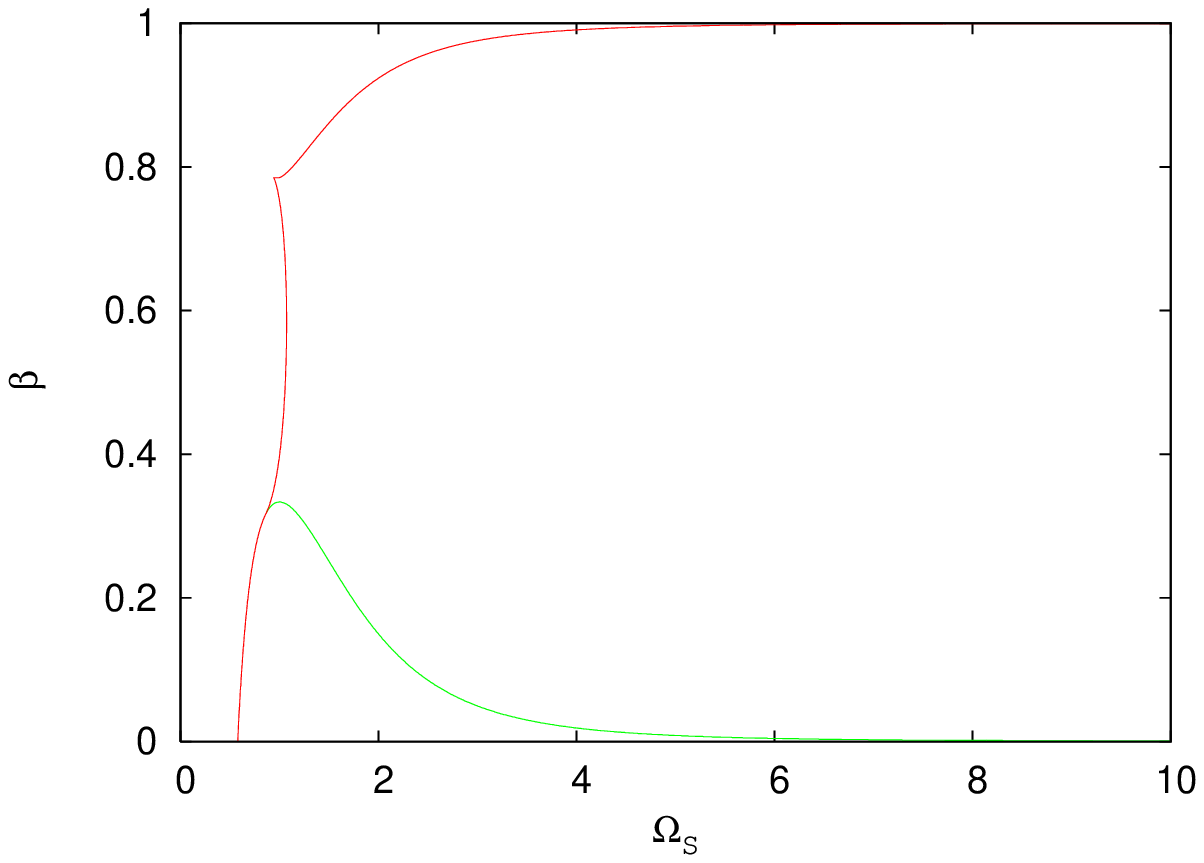}
\includegraphics[width=0.32\textwidth]{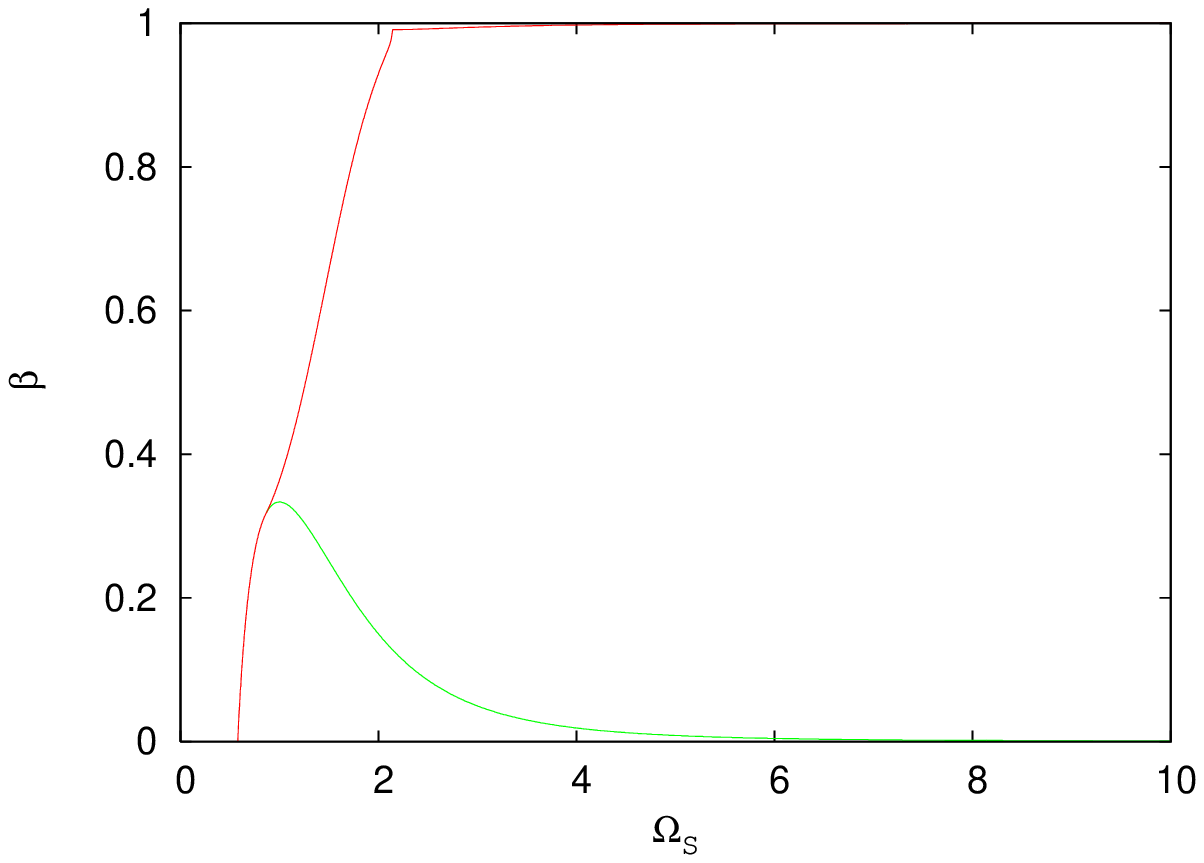}
\includegraphics[width=0.32\textwidth]{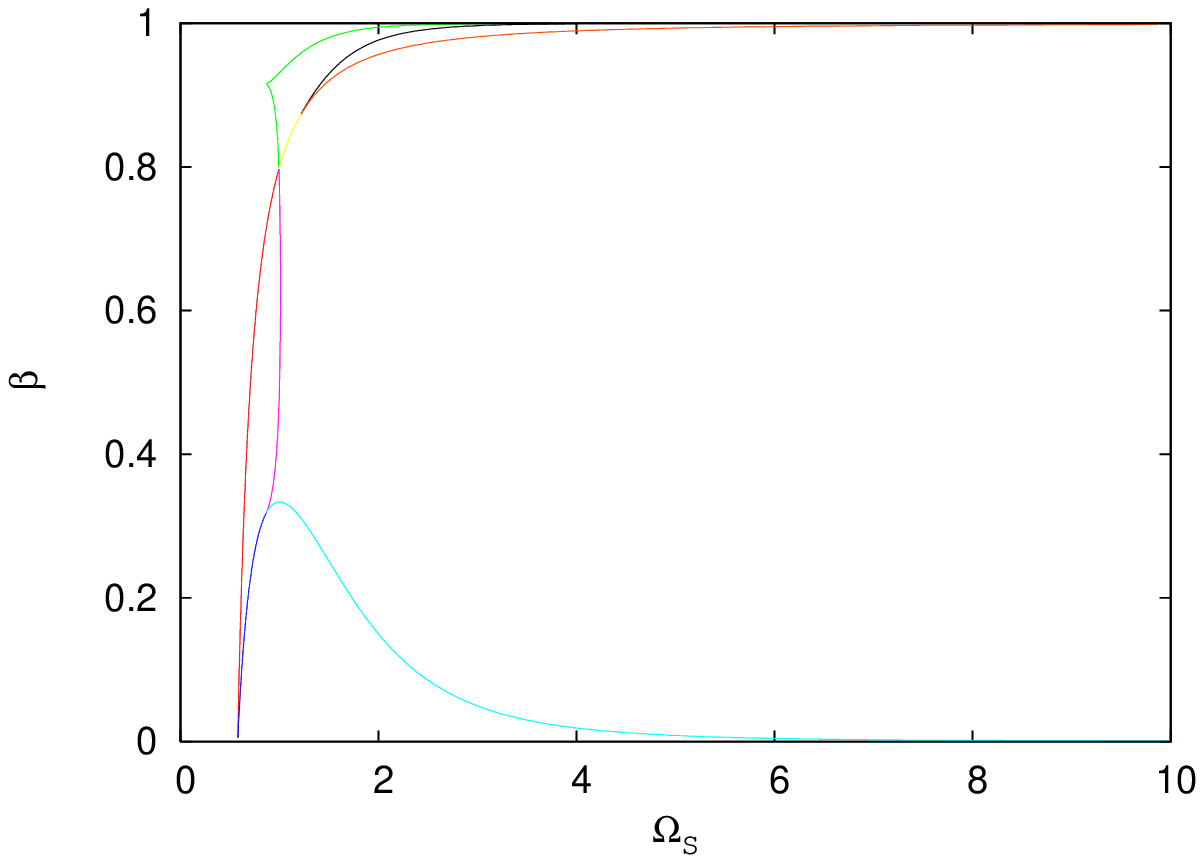}
\caption{Left panel: 
The boundary of the set $C_{\rm I0}$ (red) for the PSD (\ref{RescaledDoppler}). 
Parameters inside of this area leads to PSD with a more than one inflection point. Unlike the power spectrum (\ref{2Rorentziany})
$S_{\rm rD}(\omega)$ can have a global maximum at $\omega\approx\Omega_{\rm s}$. In some, $C_{\rm I1}$ (middle) and $C_{\rm Iq}$ (right) 
of the power spectrum (\ref{RescaledDoppler}). The common feature of in the three panels is the critical curve (\ref{BetaCrit}).}
\label{InflexLorDop}
\end{figure*}

\label{secDoppler}
In analogy with the previous section, we start the investigation with
the simplest case of identical exponential flares modulated by sine function,
\begin{equation}
I(t,\tau,\Omega_{\rm s})=I_0\,\Big(A+B\sin\big(\Omega_{\rm s} t+C\big)\Big)\;\mbox{e}^{-t/\tau}\,\theta(t).
\end{equation} 
Without loss of generality we can put $\tau=1$ and rescale the power spectrum to the form
\begin{equation}
S_{\rm rD}(\omega)=\frac{\beta}{1+\omega^2}+\frac{1-\beta}{1+(\omega-\Omega_{\rm s})^2}
+\frac{1-\beta}{1+(\omega+\Omega_{\rm s})^2},
\label{RescaledDoppler}
\end{equation}
where $\beta$ is taken from the interval $\langle 0,1\rangle$. 
Figure \ref{MaxLorDopp} shows the boundary of set of the doubly peaked $\omega\, S_{\rm rD}(\omega)$
in the parameter space and the break time-scales $\tau_{\rm M}$ for some choices of $\beta$
and $\Omega_{\rm s}$.
The structure of the set $C_{\rm M}$ corresponds again to the cusp catastrophe. However,
the structure inflection points, shown in the figure \ref{InflexLorDop}, is more rich.

Unmodulated PSDs of the form (\ref{SimSpec}) always satisfy the conditions 
$S''(\omega)>0$ for $\omega\rightarrow\infty$ and $S''(0)=-2{\rm E}[\tau^2]<0$, and hence they have an odd number 
of inflection points. However, the power spectrum (\ref{RescaledDoppler}) can be convex both at 
zero and in the limit of high frequencies
and therefore it can have an even number of inflection points. These two cases 
are separated by a curve $\beta_{\rm crit}(\Omega_{\rm s})$ given by the 
condition $S_{\rm rD}''(0)=0$. By direct
calculation we find
\begin{equation}
\beta_{\rm crit}(\Omega_{\rm s})=2\frac{3\Omega_{\rm s}^2-1}
{\Omega_{\rm s}^6+3\Omega_{\rm s}^4+9\Omega_{\rm s}^2-1}.
\label{BetaCrit}
\end{equation}

\subsection{General periodic modulation}
\label{GenPeriodMod}
We assumed that the exponentially decaying flares are at the same time orbiting in 
the equatorial plane of an accretion disc and that the observed signal is periodically modulated
by the Doppler effect and abberation. It has been shown (Paper~I) that 
the resulting power spectrum is described by the formula
\begin{equation}
S(\omega)=\sum\limits_{n=-\infty}^\infty\int\limits_{R_{\rm in}}^{R_{\rm out}}
\int\limits_{\tau_{\rm min}}^{\tau_{\rm max}}
\frac{\lambda I_0^2(\tau)\,\tau^2\left|c_n(r)\right|^2}
{1+\tau^2\left(\omega-n\Omega(r)\right)^2}
\,p(\tau,r)\,{\rm d}\tau{\rm d}r,
\label{DiskSpec}
\end{equation}
where $\Omega(r)$ is the Keplerian orbital frequency, $c_n(r)$ are the Fourier coefficients
of the periodic modulation, and $p(\tau,r)$ is the probability density of finding a flare with
the lifetime $\tau$ at the radius $r$. The formula (\ref{DiskSpec}) is far too complicated for  
calculations. Apparently, the factor $I_0^2(\tau)\,\tau^2$ can be
again absorbed into the probability distribution. Integration over $r$ and 
summation over $n$ can be transformed into a single integral by appropriate change of the variables, 
$r\rightarrow \Omega(r)/n$. Finally, we can rewrite the equation (\ref{DiskSpec}) as
\begin{eqnarray}
S(\omega)&=&\int\limits_{-\infty}^\infty\int\limits_{\tau_{\rm min}}^{\tau_{\rm max}}
\frac{1}{1+\tau^2\left(\omega-\Omega\right)^2}\,\tilde{p}(\tau,\Omega)\,{\rm d}\tau{\rm d}\Omega
\nonumber\\&=&{\rm E}\left[\frac{1}{1+\tau^2\left(\omega-\Omega\right)^2}\right].
\label{GenSpecDopp}
\end{eqnarray}
The function $\tilde{p}(\tau,\Omega)$ contains contributions from all harmonics of the orbital
frequencies $\Omega(r)$ and is by construction symmetrical in the second variable
$\tilde{p}(\tau,\Omega)=\tilde{p}(\tau,-\Omega)$. The distribution
\begin{eqnarray}
 \tilde{p}(\tau,\Omega)
&=&\delta(\tau-1)\Big[\beta\delta(\Omega)\nonumber\\ &+&(1-\beta)(\delta(\Omega-\Omega_{\rm s})
+\delta(\Omega+\Omega_{\rm s}))\Big]
\end{eqnarray}
leads to the PSD (\ref{RescaledDoppler}).

\begin{figure*}[tbh]
\begin{center}
\includegraphics[width=0.33\textwidth]{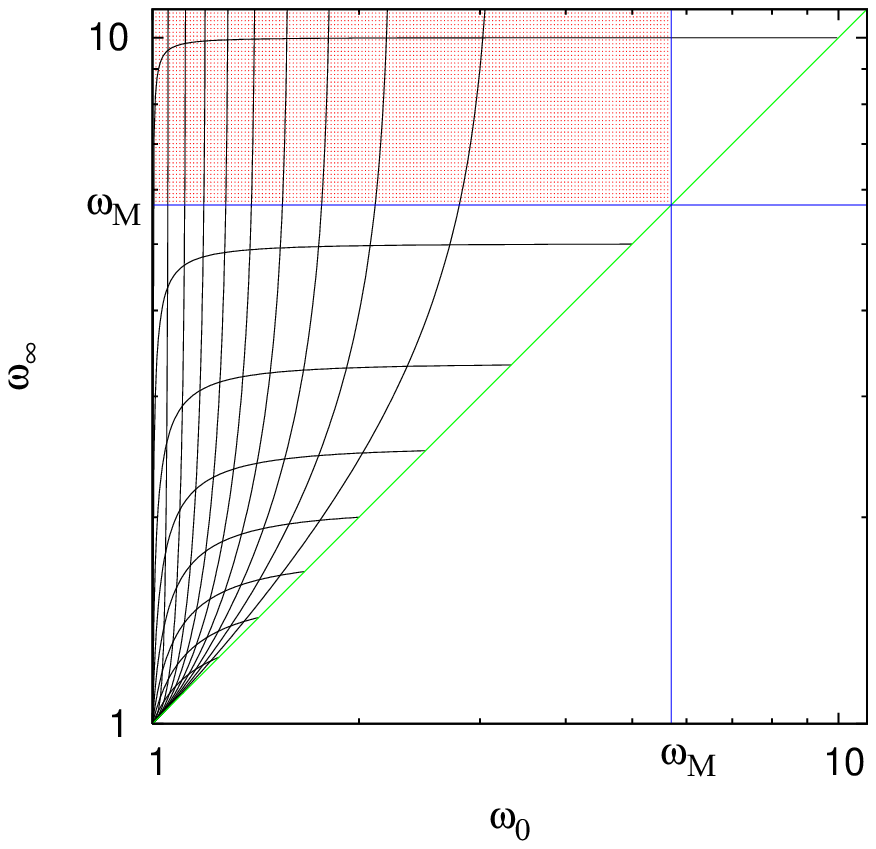}
\includegraphics[width=0.33\textwidth]{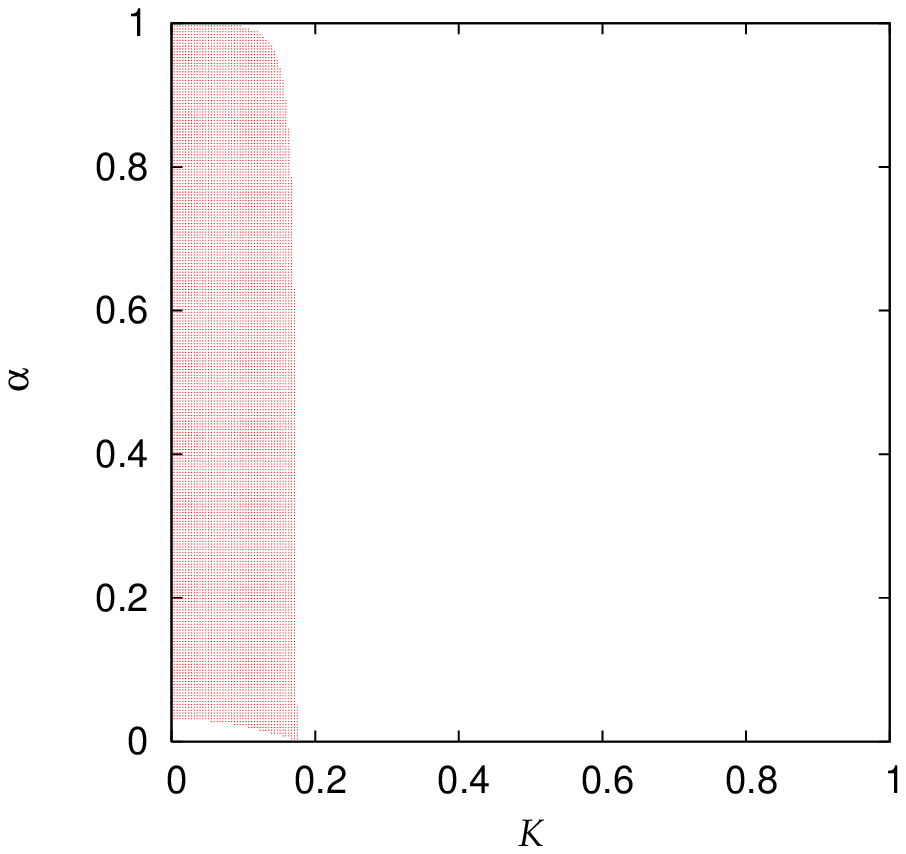}
\includegraphics[width=0.33\textwidth]{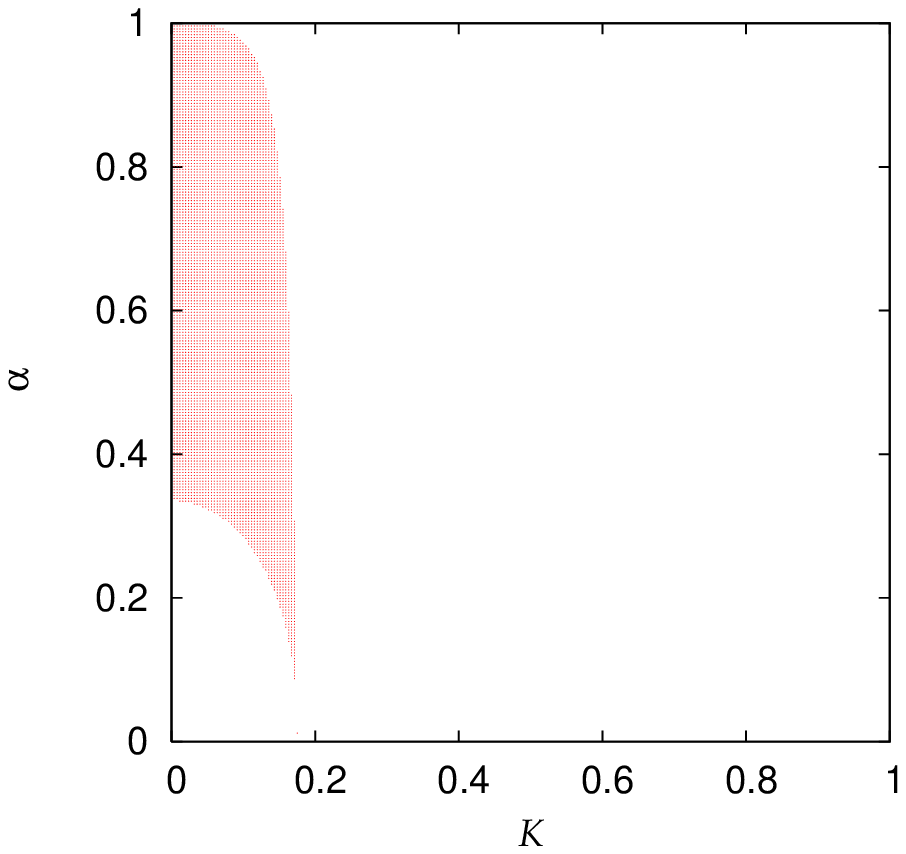}
\end{center}
\caption{
Left panel: Graph of $\omega_0=\sqrt{g(0)}$ versus $\omega_\infty=\sqrt{g(\infty)}$ calculated for the PSD from eq.\ (\ref{2Rorentziany}). 
The deformed grid of contour lines corresponds to curves of $\alpha={\rm const}$ and $K={\rm const}$ in the parameter space of the model, 
where $\omega_\infty\geq\omega_0$ by construction. It follows that the image of the entire parameter space lies above the green diagonal 
line of $\omega_\infty=\omega_0$. 
Middle panel: The upper-left rectangular sector of acceptable pairs of $\omega_0$ and $\omega_\infty$
from the left panel is projected back onto the ($\alpha,K$) parameter space. Assuming the first-order boundaries of $\omega_0=\sqrt{g(0)}$ 
and $\omega_\infty=\sqrt{g(\infty)}$, the pairs of $\alpha$ vs.\ $K$ outside of the filled (red) area lead to the 
power spectrum, which can {\em{}not} have
any local extrema at $\omega_{\rm M}$. These parameter values are therefore ruled out. Right panel: Analogical to the middle
panel, but employing tighter (second-order) boundaries $\omega_0=\sqrt{g^{(2)}(0)}$ and $\omega_\infty=\sqrt{g^{(2)}(\infty)}$;
in this case, the parameter constraints are more stringent (the filled area is smaller).}
\label{ParamConst}
\end{figure*}

We now repeat the procedures which, applied to equation (\ref{SimSpec}), 
led to the definition of the functions $g$ and $h$. We begin with the local extrema
of $\omega S(\omega)$. By differentiating (\ref{GenSpecDopp}), we find
\begin{equation}
\frac{{\rm d}}{{\rm d}\omega}[\omega S(\omega)]=
{\rm E}\left[\frac{1+\tau^2\Omega^2-\tau^2\omega^2}
{\left(1+\tau^2\left(\omega-\Omega\right)^2\right)^2}\right].
\end{equation}
The term $1+\tau^2\Omega^2$ is always positive. Therefore, we can in analogy to eq. (\ref{g-def})
define the function $g_\Omega(x)$ as
\begin{equation}
g_\Omega(\omega^2)=\frac{{\rm E}\left[(1+\tau^2\Omega^2)\left(1+\tau^2(\omega-\Omega)^2\right)^{-2}\right]}
{{\rm E}\left[\tau^2\left(1+\tau^2(\omega-\Omega)^2\right)^{-2}\right]},
\end{equation}
and observe that the position of the local extrema $\omega^2_{\rm M}=x_{\rm M}$ is again given
by the fixed point
\begin{equation}
g_\Omega(x_{\rm M})=x_{\rm M}.
\end{equation}
The function $g_\Omega$ is positive and both $g_\Omega(0)$ and $g_\Omega(\infty)$ are finite.
Unfortunately, in a general case, $g_\Omega(x)$ is not a non-decreasing function. To be able to use the techniques
developed in section \ref{sec1} we have to define a perturbed version of $g_\Omega(x)$ analogical 
to (\ref{gf-def}) as

\begin{equation}
g_{_{[f]}\Omega}(\omega^2)=\frac{{\rm E}\left[(1+\tau^2\Omega^2 +\omega^2 f(\omega^2|\tau,\Omega))
L^2(\omega|\tau,\Omega)\right]}
{{\rm E}\left[(\tau^2+f(\omega^2|\tau,\Omega))L^2(\omega|\tau,\Omega)\right]},
\label{gfOmega-def}
\end{equation}
where $L(\omega|\tau,\Omega)=\left(1+\tau^2(\omega-\Omega)^2\right)^{-1}$ is  the elementary Lorentzian term.
The perturbation $f(\omega^2|\tau,\Omega)$ has to be finite at $\omega=0$ and to decay faster than $\omega^{-1}$
to ensure the finiteness of $g_{_{[f]}\Omega}(0)$ and $g_{_{[f]}\Omega}(\infty)$. 
Again, the positivity of the derivative of $g_{_{[f]}\Omega}$ has to be tested for every particular 
choice $f$. For the power spectrum (\ref{RescaledDoppler}) it can be shown that 
\begin{equation}
f(\omega^2|\tau,\Omega)=\frac{\Omega_{\rm s}^3+1}{1+\omega^2}
\end{equation}
leads to a non-decreasing and positive function $g_{_{[f]}\Omega}(\omega^2)$.

A general calculation of the points of inflection is even more problematic. One can
proceed in the following way.
Analogically to equations (\ref{HGamma}) and (\ref{HGammaSolv}), the function $h_{\Omega}(\omega)$
is given by the solution of the following equations:
\begin{eqnarray}
3{\rm E}\left[\tau^4 L^3(\omega_{\rm I0}|\tau,\Omega)\right]\omega_{\rm I0}^3
-6{\rm E}\left[\tau^2\Omega L^3(\omega_{\rm I1}|\tau,\Omega)\right]\omega_{\rm I0}\nonumber\\
+{\rm E}\left[\tau^2(3\tau^2\Omega^2-1) L^3(\omega_{\rm I0}|\tau,\Omega)\right]=0,\\
{\rm E}\left[\tau^4 L^3(\omega_{\rm I1}|\tau,\Omega)\right]\omega_{\rm I1}^3
-3{\rm E}\left[\tau^2(1+\tau^2\Omega^2) L^3(\omega_{\rm I1}|\tau,\Omega)\right]\omega_{\rm I1}\nonumber\\
+2{\rm E}\left[ \tau^2(1+\tau^2\Omega^2)L^3(\omega_{\rm I1}|\tau,\Omega)\right]=0.
\end{eqnarray}
The positivity of the derivative of $h_\Omega$ has to be ensured by the 
perturbation procedure.

\section{Discussion and Conclusions}
\label{conclusions}
We have described a general formalism that can be used to characterize
the overall form of a power spectrum, namely, we determined the
constraints on the functional form of the PSD shape that follow from the
occurrence of local peaks. In particular, our approach allowed us to
distinguish between PSDs that have the form of multiple power-law
profiles with different numbers of local maxima.
Our methodology is useful in the context of non-monotonic PSD profiles.
In fact, Figs. \ref{zobak} and \ref{kridlo} demonstrate that
the occurrence of separate peaks of the PSD can put tight constraints
on the model. The correct interpretation of these strict constraints still remains
to be found. Naturally, a possible line of interpretation suggests that the
spot/flare scenario is not generic and needs some kind of fine-tuning of
the model parameters to reproduce observations. However, this
calls for a more thorough investigation; in the present paper we only explored
a highly idealized scheme in which strong assumptions were imposed.
More complicated models (for instance involving avalanches) also exhibit
a more varied behaviour.

A potential application of our approach is illustrated by Figure \ref{ParamConst}, where we construct a mesh 
of two-dimensional contour-lines in $(\alpha,K)$ parameter space of the model. 
Assuming that the exemplary power spectrum has a local maximum (or minimum) at frequency 
$\omega_{\rm M}$, it follows directly from eq.\ (\ref{GenaralGIneq}) that the 
three frequencies must satisfy the relation $\omega_0\leq\omega_{\rm M}\leq\omega_\infty$. 
Only those pairs of parameters $\alpha$ and $K$ whose images are within 
the upper-left sector (Fig.\ \ref{ParamConst}, left panel), as described by the 
mentioned inequality, can produce a power spectrum with the required feature 
at $\omega=\omega_{\rm M}$.

An example for this is Cyg X-1, which has been described as a
superposition of several Lorentzians in the X-ray band light curve
\citep{2003A&A...407.1039P}. These authors have demonstrated that
the change of the timing parameters of the source is mainly caused by a
strong decrease in the amplitude of one of the Lorentzian components
present in the PSD. This characteristic is closely connected with
the normalization of the model components. During the state transition
of the source one of the Lorentzians becomes suppressed relative to the
other. It was suggested that this behaviour is associated with the
accretion disc corona, which is believed to be responsible for the hard-state spectrum.
The evolution of the PSD form can be studied in terms of our method,
which allowed us to discuss the entire class of multiple-power-law PSDs
within a uniform systematic scheme. Although it is a versatile
approach, its practical use can be illustrated in a very simple way. 

The hard-state X-ray spectrum is very likely related with the presence
of soft photons from the accretion disc, which are Compton up-scattered
in a hot electron gas. The state transitions are then caused by the disappearance 
of the Comptonizing medium. Such a change in the coronal configuration could 
explain the change of the mutual normalization of the Lorentzian components, 
namely, the disappearance or recurrence of local peaks in the PSD.
In the idea of coronal flares modulating the accretion disc variability 
of the source, which seems to be relevant for Cyg X-1, the position of the 
source in Fig.\ \ref{ParamConst} will constrain possible parameter 
values of the model.

Finally, we have checked that a certain proportionality between the light curve rms
and the flux does exist in our model as well. This is an interesting fact by itself, 
however, it is not clear at present whether the slope and the scatter of the rms-flux 
relation are in reasonable agreement with the
actual data, and how generic the model is. Furthermore detailed investigations 
are needed to clarify whether the spot model could satisfy the constraints 
arising beyond the PSD profiles. Moreover, it remains to be seen if our model
requires some kind of special fine-tuning of the parameters (such as the
inclination angle), which would make this explanation less likely (work in 
progress).

\begin{acknowledgements} 
The research leading to these results has received funding from the Czech Science 
Foundation and Deutsche Forschungsgemeinschaft collaboration project (VK, GACR-DFG 13-00070J). 
We also acknowledge the Polish grant NN 203 380136 (BC) and the French GdR PCHE (RG). 
Part of the work was supported by the European Union Seventh Framework Programme 
under the grant agreement No.\ 312789 (MD, BC, RG). The
Astronomical Institute has been operated under the program RVO:67985815 in
the Czech Republic (TP).
\end{acknowledgements}

\bibliographystyle{aa} 
\bibliography{9999} 

\appendix
\section{Monotonicity of the functions {\fontshape{it}\selectfont{}g} and {\fontshape{it}\selectfont{}h}}
\label{appa}
We prove that the derivatives of $g_k(x)$ and $h_k(x)$ are non-negative. 
We define an operator ${\rm F}_{x,k}[.]$ by its action on an arbitrary function $f(\tau)$ as
\begin{equation}
{\rm F}_{x,k}[f(\tau)]=N_{x,k}\,{\rm E}\left[\frac{f(\tau)}{\left(1+\tau^2\omega^2\right)^{k+3}}\right],
\end{equation}
where $N_{x,k}$ is a normalization constant ensuring that ${\rm F}_{x,k}[1]=1$ for all $k$ and $x$.
For a fixed value of $x$ it acts as an mean value operator with a probability distribution 
$p_x(\tau)=p(\tau)/\left(1+\tau^2\omega^2\right)^{-k-3}$. Using this operator, we can express
the derivatives as
\begin{eqnarray}
\frac{{\rm d}g_k(x)}{{\rm d}x}&=&(k+2)\frac{{\rm F}_{x,k}[\tau^4]-\left({\rm F}_{x,k}[\tau^2]\right)^2}
{\left({\rm F}_{x,k}[\tau^2(1+\tau^2x)]\right)^2},\label{gDeriv}\\
\frac{{\rm d}h_k(x)}{{\rm d}x}&=&(k+3)\frac{{\rm F}_{x,k+1}[\tau^6]{\rm F}_{x,k+1}[\tau^2]-\left({\rm F}_{x,k+1}[\tau^4]\right)^2}
{\left({\rm F}_{x,k+1}[\tau^4(1+\tau^2x)]\right)^2}.\label{hDeriv}
\end{eqnarray}
The denominators in both equations are always non-negative. Furthermore, non-negativity of the numerator of (\ref{gDeriv}) follows
from the Jenssen theorem. The numerator of (\ref{hDeriv}) is a determinant of a matrix of the following
quadratic form:
\begin{eqnarray}
\left(\begin{array}{c}
a\\b
\end{array}\right)^{\rm T}
\left(\begin{array}{cc}
{\rm F}_{x,k+1}[\tau^6] & {\rm F}_{x,k+1}[\tau^4]\\
{\rm F}_{x,k+1}[\tau^4] & {\rm F}_{x,k+1}[\tau^2]
\end{array}\right)
\left(\begin{array}{c}
a\\b
\end{array}\right) & \nonumber \\
& \hspace*{-3em} ={\rm F}_{x,k+1}[(a\tau^3+b\tau)^2]\geq 0.
\end{eqnarray}
Because the form is positively (semi-)definite, the determinant must be a non-negative function.

\section{Operator associated with the polynomial {\fontshape{it}\selectfont{}Q}}
We give an explicit formula for the operator $\mathbf{\hat{Q}}_k$.
We define operators $\mathbf{\hat{N}}$ and $\mathbf{\hat{D}}$ as
\begin{equation}
\mathbf{\hat{N}}=\frac{1}{\tau}\frac{{\rm d}}{{\rm d}\tau},\quad
\mathbf{\hat{D}}=\tau^3\frac{{\rm d}}{{\rm d}\tau}.
\end{equation}
It can be proven by direct calculation that
\begin{eqnarray}
\mathbf{\hat{N}}^l\frac{1}{1+\tau^2\omega^2}\,=\frac{l!(-2)^l\omega^{2l}}{\left(1+\tau^2\omega^2\right)^{l+1}},\\
\mathbf{\hat{D}}^l\frac{\tau^2}{1+\tau^2\omega^2}\,=\frac{l!(2)^l\omega^{2l}\tau^{2l+1}}{\left(1+\tau^2\omega^2\right)^{l+1}}.
\end{eqnarray}
From these relations it follows, for arbitrary $l$ and $m$,
\begin{eqnarray}
\frac{(-1)^l}{(l+m)!2^{l+m}}&\tau^{-2m-1}&
\mathbf{\hat{D}}^m\,\tau^{2l+2}\,\mathbf{\hat{N}}^l\,\frac{1}{1+\tau^2\omega^2}\nonumber\\
&=&\frac{\left(\tau^2\omega^2\right)^l}{\left(1+\tau^2\omega^2\right)^{l+m+1}}.
\end{eqnarray}
Assuming that $Q_k(z)=\sum_{l=0}^k q_l\, z^l$, we can define the operator $\mathbf{\hat{Q}}_k$ as
\begin{equation}
\mathbf{\hat{Q}}_k=\frac{1}{k!2^k}\left(\frac{1}{\tau}\right)^{2k+1}
\sum\limits_{l=0}^k (-1)^l\, q_l\,\tau^l\, \mathbf{\hat{D}}^{k-l}\,\tau^{2l+2}\,\mathbf{\hat{N}}^l.
\label{QkDef}
\end{equation}
The polynomials  $Q_k$ and $P_k$ are mutually related by the formula 
\begin{equation}
Q_k(\tau^2\omega^2)=(1+\tau^2\omega^2)^{k+1}\left|\sum\limits_{l=0}^k l!\,p_l\,(1+i\tau\omega)^{-l-1}\right|^2,
\end{equation} 
where $p_l$ are the coefficients of $P_k$.
\end{document}